\documentclass[useAMS,usenatbib]{mn2e}
\usepackage{amssymb,amsmath}
\usepackage[normalem]{ulem}
\usepackage{times}
\usepackage{ctable,longtable}
\usepackage{graphicx}
\usepackage[colorlinks=true, linkcolor=blue, citecolor=blue, urlcolor=blue]{hyperref}
\usepackage{longtable}
\usepackage{lscape}
\usepackage[varg]{txfonts}
\usepackage{etoolbox}
\makeatletter
\newcount\c@additionalboxlevel
\newcount\c@maxboxlevel
\patchcmd\@combinedblfloats{\box\@outputbox}{%
  \stepcounter{additionalboxlevel}%
  \box\@outputbox
}{}{\errmessage{\noexpand\@combinedblfloats could not be patched}}

\AtBeginShipout{%
  \ifnum\value{additionalboxlevel}>\value{maxboxlevel}%
    \typeout{Warning: maxboxlevel might be too small, increase to %
      \the\value{additionalboxlevel}%
    }%
  \fi 
  \@whilenum\value{additionalboxlevel}<\value{maxboxlevel}\do{%
    \typeout{* Additional boxing of page `\thepage'}%
    \setbox\AtBeginShipoutBox=\hbox{\copy\AtBeginShipoutBox}%
    \stepcounter{additionalboxlevel}%
  }%
  \setcounter{additionalboxlevel}{0}%
}
\makeatother

\title[The ISM in Andromeda's dSphs: I. Interstellar dust]{The interstellar medium in Andromeda's dwarf spheroidal galaxies: I. Content and origin of the interstellar dust} 
\author[De Looze et al.]{Ilse De Looze$^{1,2,3}$\thanks{E-mail: idelooze@star.ucl.ac.uk, ilse.delooze@ugent.be}, Maarten Baes$^{2}$, George J. Bendo$^{4}$, Jacopo Fritz$^{2,5}$, M{\'e}d{\'e}ric Boquien$^{3,6}$, 
\newauthor Diane Cormier$^{7}$, Gianfranco Gentile$^{2,8}$, Robert C. Kennicutt$^{3}$, Suzanne C. Madden$^{9}$, 
\newauthor Matthew W.~L. Smith$^{10}$, Lisa Young$^{11,12}$  \\
$^{1}$ Department of Physics and Astronomy, University College London, Gower Street, London WC1E 6BT, UK\\
$^{2}$ Sterrenkundig Observatorium, Universiteit Gent, Krijgslaan 281 S9, B-9000 Gent, Belgium \\
$^{3}$ Institute of Astronomy, University of Cambridge, Madingley Road, Cambridge, CB3 0HA, UK \\
$^{4}$ UK ALMA Regional Centre Node, Jodrell Bank Centre for Astrophysics, School of Physics and Astronomy, \\ University of Manchester, Oxford Road, Manchester M13 9PL, United Kingdom \\
$^{5}$ Instituto de Radioastronom{\'i}a y Astrof{\'i}sica, IRyA, UNAM, Campus Morelia, A.P. 3-72, C.P. 58089 Michoac{\'a}n, Mexico \\
$^{6}$ Unidad de Astronom{\'i}a, Fac. Cs. B{\'a}sicas, Universidad de Antofagasta, Avda. U. de Antofagasta 02800, Antofagasta, Chile\\ 
$^{7}$ Zentrum f{\"u}r Astronomie der Universit{\"a}t Heidelberg, Institut f{\"u}r Theoretische Astrophysik, Albert-Ueberle Str. 2, D-69120 Heidelberg, Germany \\ 
$^{8}$ Astrophysical Institute, Vrije Universiteit Brussel, Pleinlaan 2, 1050 Brussels, Belgium \\
$^{9}$ Laboratoire AIM, CEA, Universit{\'e} Paris VII, IRFU/Service d$'$Astrophysique, Bat. 709, 91191 Gif-sur-Yvette, France \\
$^{10}$ School of Physics and Astronomy, Cardiff University, Queens Buildings, The Parade, Cardiff CF24 3A A \\
$^{11}$ Physics Department, New Mexico Institute of Mining and Technology, Socorro, NM 87801, USA \\
$^{12}$ Academia Sinica Institute of Astronomy \& Astrophysics, PO Box 23-141, Taipei 10617, Taiwan, R.O.C. }
\begin{document}

\date{Accepted 2016, April 8; Received 2016, February 20}

\pagerange{\pageref{firstpage}--\pageref{lastpage}} \pubyear{2016}

\maketitle

\label{firstpage}

\begin{abstract}
Dwarf spheroidal galaxies are among the most numerous galaxy population in the Universe, but their main formation and evolution channels are still not well understood. The three dwarf spheroidal satellites (NGC\,147, NGC\,185, and NGC\,205) of the Andromeda galaxy are characterised by very different interstellar medium (ISM) properties, which might suggest them being at different galaxy evolutionary stages. While the dust content of NGC\,205 has been studied in detail by \citet{2012MNRAS.423.2359D}, we present new \textit{Herschel} dust continuum observations of NGC\,147 and NGC\,185.

The non-detection of NGC\,147 in \textit{Herschel} SPIRE maps puts a strong constraint on its dust mass ($\leq$ 128$^{+124}_{-68}$ M$_{\odot}$). For NGC\,185, we derive a total dust mass $M_{\text{d}}$ = 5.1$\pm$1.0 $\times$ 10$^{3}$ M$_{\odot}$, which is a factor of $\sim$ 2-3 higher than that derived from \textit{ISO} and \textit{Spitzer} observations and confirms the need for longer wavelength observations to trace more massive cold dust reservoirs. 

We, furthermore, estimate the dust production by asymptotic giant branch (AGB) stars and supernovae (SNe). For NGC\,147, the upper limit on the dust mass is consistent with expectations of the material injected by the evolved stellar population. In NGC\,185 and NGC\,205, the observed dust content is one order of magnitude higher compared to the estimated dust production by AGBs and SNe. Efficient grain growth, and potentially longer dust survival times (3-6 Gyr) are required to account for their current dust content. Our study confirms the importance of grain growth in the gas phase to account for the current dust reservoir in galaxies. 

\end{abstract}

\begin{keywords}
ISM: dust -- galaxies: dwarf -- galaxies: individual: NGC 147 -- galaxies: individual: NGC 185 -- galaxies: individual: NGC 205 -- Local Group -- infrared: ISM
\end{keywords}

\section{Introduction}

Dwarf spheroidal galaxies (dSph) belong to the most numerous galaxy population in the Universe, but their low surface brightness hampers their detection. 
The Local Group offers the nearest laboratory to analyse the formation and evolution processes of the low-luminosity dSph galaxy population.
The historical picture of dSphs containing relatively little interstellar material has changed during the past few years owing to the improved spatial resolution and sensitivity of infrared, (sub-)millimeter and radio observing facilities (e.g., \textit{Herschel}, EVLA, BIMA, IRAM).
The detection of gas (e.g., \citealt{1996ApJ...470..781W,1996ApJ...464L..59Y,1997ApJ...476..127Y,2001AJ....122.1747Y,2007A&A...474..851D,2009A&A...498..407G}) and dust (e.g., \citealt{2010A&A...518L..54D,2013A&A...552A...8D,2013MNRAS.436.1057D}) shows that a subpopulation of dSphs does contain a significant reservoir of interstellar material. The presence of a young stellar population, within galaxies otherwise dominated by old stars, also demonstrates that some dSphs have experienced a recent episode of star formation (e.g., \citealt{1993A&A...271...51P,2005ApJ...629L..29B}). 

In this paper, we focus on the study of the dust reservoir in the dSph companions of the Andromeda galaxy (M\,31). The dust content of the dSph galaxy NGC\,205 was analysed in \citet{2012MNRAS.423.2359D}. This paper focuses on the two remaining spheroidal dwarfs, NGC\,147 and NGC\,185. The dust reservoir in these two galaxies has been observed and analysed in the past by IRAS \citep{1988ApJS...68...91R}, ISO \citep{2004ApJS..151..237T} and \textit{Spitzer} \citep{2010ApJ...713..992M}.
Overall, the dSph galaxies NGC\,147 and NGC\,185 appear very similar in terms of their galaxy mass \citep{2010ApJ...711..361G}, stellar kinematics \citep{2010ApJ...711..361G} and C star fraction \citep{2005AJ....130.2087D}\footnote{Deep imaging from the Pan-Andromeda Archaeological Survey (PAndAS) recently revealed that the effective radius in NGC\,147 is about three times larger compared to NGC\,147 and that NGC\,185 is about one magnitude brighter compared to NGC\,185 \citep{2014MNRAS.445.3862C}.}. From a detailed analysis of their star formation history, \citet{2015ApJ...811..114G}, however, concluded that the bulk of stellar mass was formed around 12.5 Gyr ago in NGC\,185, while most of the stars in NGC\,147 only formed about 5 to 7 Gyr ago. The latter result is interpreted as an earlier infall time for NGC\,185 compared to NGC\,147 into the M\,31 group. Other than the age variations for the bulk of stars in the two dwarf spheroidals, NGC\,185 is characterised by a more recent star formation episode limited to its central regions, that started a few 100 Myr ago. A similar central concentration of young stars is also found in the other Local Group dSph NGC\,205, suggesting that the build-up of mass from evolved stars and planetary nebulae has initiated a recent cycle of star formation in their central regions. The ongoing star formation in NGC\,185 contrasts with the lack of any sign of recent star formation activity during the last 1 Gyr in NGC\,147 \citep{1997AJ....113.1001H} and the absence of any detected ISM material in NGC\,147 \citep{1998ApJ...499..209W}.

The variation in star formation history and ISM content observed in the three dwarf spheroidal companion galaxies of M\,31 seems to imply that the dwarfs are at a different stage of evolution and/or might have experienced different levels of interaction with the group environment. In a companion paper \citep{DeLooze_paper2}, we revise the gas content in the three dSphs and discuss their gas content, ISM properties and gas-to-dust ratio in view of galaxy evolution processes. In this work, we compare the dust mass content derived from \textit{Herschel} observations in the three dwarf spheroidal galaxies with predictions of the mass return by AGB stars and supernova remnants. Based on such comparison, we aim to put constraints on the dust production and survival in the ISM of dwarf spheroidal galaxies.

Due to the proximity ($D$ $<$ 1 Mpc) of the dwarf spheroidal satellite galaxies, it is possible to identify individual AGB stars from near-infrared observations (e.g., \citealt{2004A&A...417..479B,2004A&A...418...33B,2005AJ....130.2087D,2005A&A...437...61K,2006A&A...445...69S}) and to derive realistic star formation histories based on color-magnitude diagrams (e.g., \citealt{1999AJ....118.2229M,2009A&A...502L...9M,2015ApJ...811..114G}). In our Galaxy, AGB stars have long been thought to be the main dust producers (e.g., \citealt{2005ASPC..341..605T}). In the Large Magellanic Clouds, supernovae are considered equally important as dust source compared to AGB stars \citep{2009MNRAS.396..918M}. Recent observations of core-collapse supernovae (CCSNe) at sub-millimeter (sub-mm) wavelengths (e.g., \citealt{2010A&A...518L.138B,2011Sci...333.1258M,2012ApJ...760...96G,2014ApJ...782L...2I,2015ApJ...800...50M}) and optical wavebands (e.g., \citealt{2015arXiv150906379A,2016MNRAS.456.1269B}) have predicted increased dust yields in CCSNe, which might also significantly contribute to the large dusty reservoirs observed in galaxies at an early stage of their evolution at high-redshift (e.g., \citealt{2015A&A...577A..80M}). On global galaxy scales, the total dust reservoir can, often, not be accounted for by AGB stars and SNe alone, and requires an efficient growth of dust grains in the interstellar medium (e.g., \citealt{2009MNRAS.396..918M,2012MNRAS.423...26M,2013EP&S...65..213A,2013A&A...555A..99Z,2014MNRAS.444..797M,2014A&A...563A..31R}). 

The comparison of the observed dust content in the three dwarf spheroidal companions of M\,31 with predictions of dust mass produced by AGB stars and supernova remnants allows us to put constraints on the importance of various dust production mechanisms in low-metallicity environments. 

In Section \ref{Data.sec}, we present our \textit{Herschel} observations and the ancillary dataset used in this analysis. Section \ref{Morphology.sec} discusses the detection and morphology of dust reservoirs in NGC\,147 and NGC\,185.
Section \ref{SEDmodel.sec} models the multi-waveband dust spectral energy distribution (SED) in NGC\,185 with modified black-body (MBB) and full dust models, and discusses the possible presence of a sub-millimeter excess. The origin of the dust in the dwarf spheroidal satellites of M\,31 is discussed in Section \ref{DustSource.sec}. Section \ref{Conclusions.sec} finally sums up.  

Throughout this paper, we adopt distances of $675\pm27$ kpc, $616\pm26$ kpc, and $824\pm27$ kpc to NGC\,147, NGC\,185, and NGC\,205 \citep{2005MNRAS.356..979M}, respectively. We determine the metal abundance in the ISM of NGC\,147 (0.23 Z$_{\odot}$), NGC\,185 (0.36 $Z_{\odot}$), and NGC\,205 (0.25 Z$_{\odot}$) by averaging over the oxygen abundances derived from planetary nebulae (PNe) in the centers of those galaxies (see \citealt{DeLooze_paper2} for more details). 

\section{Data}
\label{Data.sec}

\subsection{\textit{Herschel} dust continuum data}

\textit{Herschel} observations of NGC\,147 and NGC\,185 were obtained as part of the OT2 program $``$\textit{Herschel} study of the ISM in Local Group dwarf ellipticals$"$ (PI: De Looze).
The dust continuum has been mapped with the SPIRE photometer out to 2 effective radii in NGC\,147 and NGC\,185, corresponding to $6\arcmin \times 6\arcmin$ and $12\arcmin \times 12\arcmin$ maps, respectively. Each of the two areas were observed in nominal and orthogonal scan directions each with three repetitions at the medium scan speed of 20$\arcsec s^{-1}$. The SPIRE photometry observations for NGC\,185 (ObsID 1342249109) and NGC\,147 (ObsID 1342258361) were performed on August 6th 2012 and January 3rd 2013, respectively. 
All raw SPIRE data have been reduced up to level 1 in the \textit{Herschel} Interactive Processing Environment (\texttt{HIPE} v12.0. \citealt{2010ASPC..434..139O}) with calibration files v48, following the standard pipeline procedures. The level 1 data have been processed using the BRIght Galaxy ADaptive Element method (BRIGADE, \citealt{2012SmithPhD,2013MNRAS.428.1880A}, Smith et al.\,in prep.). The latter uses a custom method to remove the temperature drift and bring all bolometers to the same level (instead of the default \textit{temperatureDriftCorrection} and the residual, median baseline subtraction). Relative gain corrections were, furthermore, applied to correct for the bolometer's responsivity to extended sources. The data were then mapped using the naive mapmaking algorithm in \texttt{HIPE}. Final SPIRE maps have been obtained with pixel sizes of 6$\arcsec$, 8$\arcsec$ and 12$\arcsec$ at 250, 350 and 500\,$\mu$m, respectively. The FWHM of the SPIRE beam corresponds to 18.2$\arcsec$, 24.9$\arcsec$ and 36.3$\arcsec$ at 250, 350 and 500\,$\mu$m, respectively (see SPIRE Observers' Manual). The SPIRE maps are multiplied with $K_{\text{PtoE}}$ correction factors for conversion from point source to extended source photometric calibration. The appropriate correction factors (i.e., 91.289, 51.799, 24.039 MJy/sr (Jy beam$^{-1}$)$^{-1}$ at 250, 350 and 500\,$\mu$m, respectively; see SPIRE Observers' manual) for a constant $\nu S_{\nu}$ spectrum convert the maps from Jy beam$^{-1}$ to MJy sr$^{-1}$ units, and account for the most up-to-date measured SPIRE beam areas of 465, 823, 1769 arcsec$^2$ at 250, 350 and 500\,$\mu$m, respectively. 

For NGC\,185, SPIRE observations were complemented with PACS photometry mapping in the green (100\,$\mu$m) and red (160\,$\mu$m) photometric bands across a 8$\arcmin \times 8\arcmin$ area, with 4 repetitions in both nominal and orthogonal directions.
The PACS observations of NGC\,185 (ObsID 1342247328, 1342247329) took place on June 24th 2012.
PACS data have been processed from level 0 to level 1 in \texttt{HIPE} v12.0, using the calibration file v56. The level 1 data are processed with the map making IDL algorithm \textit{Scanamorphos} (v23, \citealt{2013PASP..125.1126R}), which is optimised to benefit from the redundancy of all coverages in every sky pixel. Final PACS maps are created with pixel sizes of 1.7$\arcsec$ and 2.85$\arcsec$ at 100\,$\mu$m and 160\,$\mu$m, respectively. The full width at half-maximum (FWHM) of the PACS beam has sizes of 6.9$\arcsec$ and 12.1$\arcsec$ at 100 and 160\,$\mu$m, respectively.

\subsection{Ancillary data}
The Infrared Array Camera (IRAC, \citealt{2004ApJS..154...10F}) and Multiband Imaging Photometer (MIPS, \citealt{2004ApJS..154...25R}) on board the \textit{Spitzer} Space Telescope observed NGC\,147 and NGC\,185 as part of the program Detailed Study of the Dust in M31's Four Elliptical Companions (PI: F. Marleau). The observing strategy and reduced IRAC and MIPS maps are presented and analysed in \citet{2010ApJ...713..992M}. We retrieved IRAC data at 3.6, 4.5, 5.8 and 8.0\,$\mu$m, and MIPS\,24, 70 and 160\,$\mu$m maps from the \textit{Spitzer} Heritage Archive\footnote{http://sha.ipac.caltech.edu/applications/Spitzer/SHA/}. The MIPS data have been reprocessed following the methods described by \citet{2012MNRAS.423..197B}. 

\subsection{Data processing}

All images are background subtracted with the background value derived as the median from a set of 50 random apertures with radius R = $4 \times \text{FWHM}$ in the field around the galaxy. 

For the dust SED fitting procedure presented in Section \ref{SEDmodel.sec}, all infrared images are convolved to match the SPIRE\,500\,$\mu$m resolution (36.3$\arcsec$) using the appropriate kernels from \citet{2011PASP..123.1218A}. All convolved images are rebinned to match the pixel grid of the SPIRE\,500\,$\mu$m frame with pixel size of 12$\arcsec$ which corresponds to a physical scale of 35.8 pc at the distance of NGC\,185. 
With the SPIRE\,500\,$\mu$m waveband providing the limiting resolution scale, we consider \textit{Spitzer} MIPS 24 and 70\,$\mu$m and \textit{Herschel} PACS 100, 160\,$\mu$m and SPIRE 250, 350, 500\,$\mu$m data to constrain the dust spectral energy distribution in NGC\,185. The MIPS\,160\,$\mu$m waveband is not considered with a resolution of 38$\arcsec$ \citep{2007PASP..119..994E}.


\section{The morphology of dust, gas and stars in NGC\,147 and NGC\,185}
\label{Morphology.sec}
\subsection{NGC\,147}
\begin{figure*}
\centering
\includegraphics[width=17.75cm]{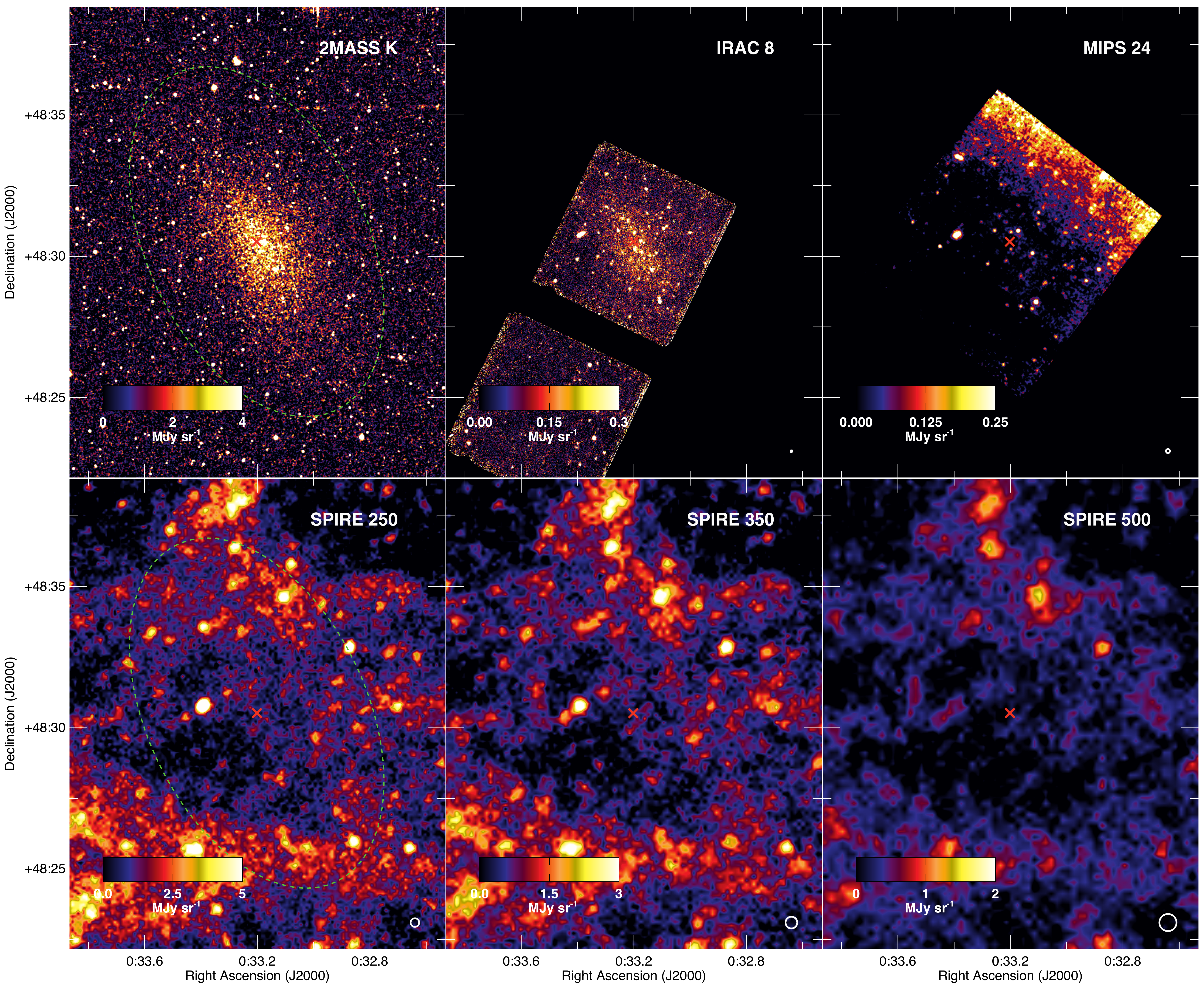}   \\
 \caption{Overview of the \textit{Spitzer} and \textit{Herschel} maps for NGC\,147. The top left panel shows a near-infrared 2MASS $K$ band image of NGC\,147 for comparison, with the ellipse indicating the extent of one-third of $B$ band 25 mag arcsec$^{-2}$ isophotal contours. The other panels, from left to right and top to bottom, show the dust emission in the IRAC\,8\,$\mu$m, MIPS\,24\,$\mu$m, and SPIRE\,250, 350 and 500\,$\mu$m wavebands, respectively. The extent of one-third of the $B$ band 25 mag arcsec$^{-2}$ isophotes is also overlaid on the SPIRE\,250\,$\mu$m image. The centre of the galaxy is indicated with a red cross. The FWHM of the PSF is indicated as a white circle in the lower right corner of each panel.}
              \label{Images_Herschel_NGC147}
\end{figure*}

\begin{figure}
\centering
\includegraphics[width=8.75cm]{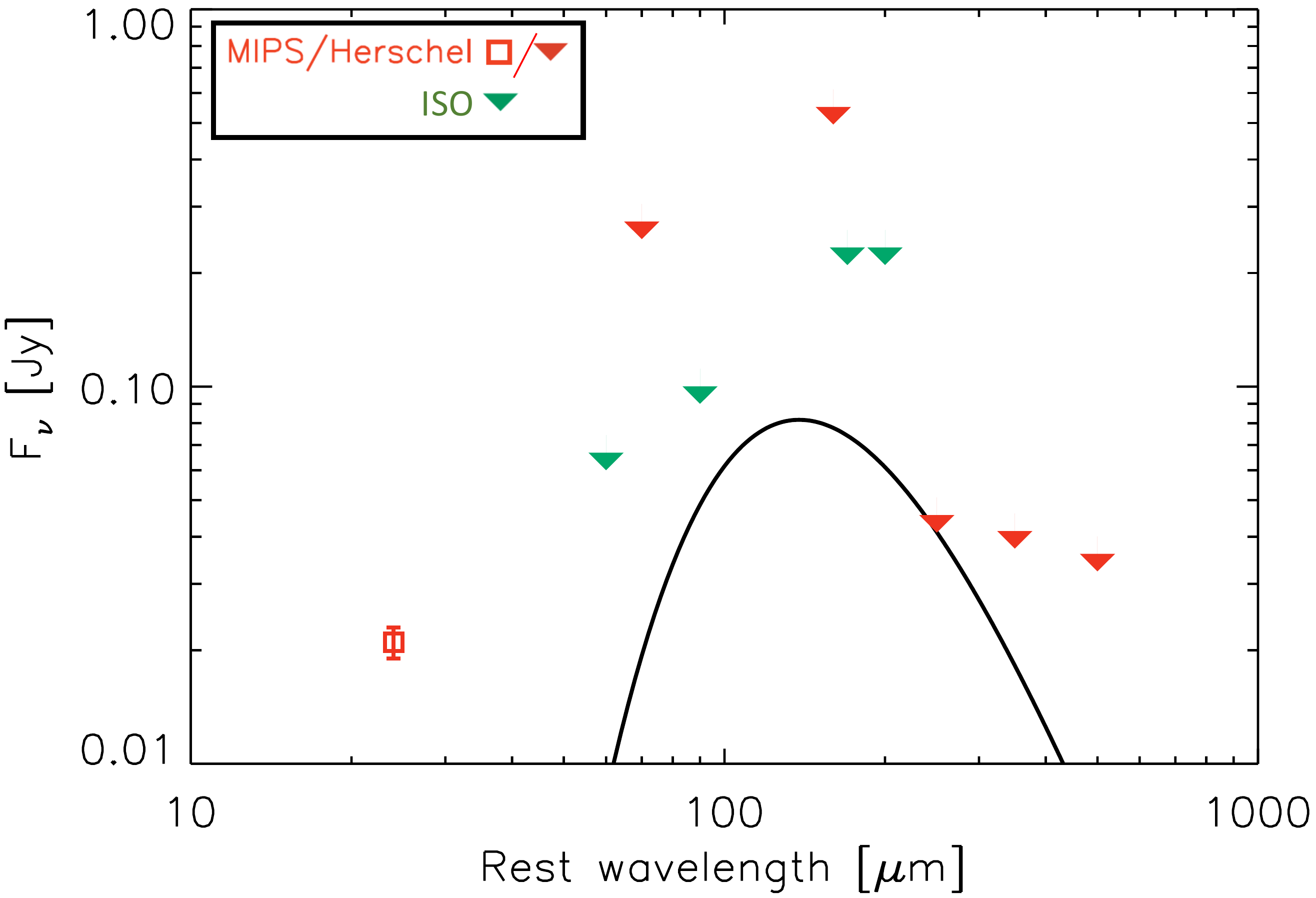}   \\
 \caption{The SED with the MIPS\,24\,$\mu$m data point (red diamond) and 3$\sigma$ upper limits (downward arrows) from \textit{Spitzer} \citep{2010ApJ...713..992M}, \textit{Herschel} (this paper) and ISO \citep{2004ApJS..151..237T} observations for NGC\,147. The modified blackbody model with emissivity index $\beta$=2 and dust temperature T$_{\text{d}}$ = 21\,K (black solid line) is scaled to the 3$\sigma$ upper limit at 250\,$\mu$m.}
              \label{ima_MBB_NGC147}
\end{figure}

Figure \ref{Images_Herschel_NGC147} shows the 2MASS $K$ band, IRAC 8\,$\mu$m, MIPS\,24\,$\mu$m, and \textit{Herschel} SPIRE photometry maps for NGC\,147. No significant sub-millimeter emission is detected within the B band 25 mag arcsec$^{-2}$ isophotal contours, but prominent Galactic cirrus emission is observed in the field around NGC\,147 (mainly towards the South-East of the galaxy). 
The background noise in the SPIRE images is determined as the standard deviation around the mean background value for 100 apertures with R=FWHM, positioned in background regions avoiding the strong cirrus emission towards the south-east of NGC\,147. We find 1$\sigma$ noise levels of 13.67, 12.40, 10.79 mJy beam$^{-1}$ in the SPIRE\,250, 350, and 500\,$\mu$m maps, respectively. We determine a 3$\sigma$ upper mass limit from the SPIRE\,250\,$\mu$m image through $M_{\text{d}}$ $\lesssim$ $\frac{S_{\text{250}} D^{2}}{\kappa_{\text{250}} B_{\text{250}}(T_{\text{d}})}$ with a dust mass absorption coefficient $\kappa_{\text{250}}$ = 0.40 m$^{2}$ kg$^{-1}$ (consistent with the \citealt{2007ApJ...657..810D} dust model), distance D and Planck function $B_{\text{250}}$ ($T_{\text{d}}$) for a dust temperature $T_{\text{d}}$. Assuming an average dust temperature $T_{\text{d}}$ $\sim$ 21 K similar to the global fitted dust temperature for NGC\,185 for a PAH+graphite+amorphous silicate dust mixture (see Section \ref{GlobalSED.sec}), we derive a 3$\sigma$ upper dust mass limit $M_{\text{d}}$ $\lesssim$ 128$^{+124}_{-68}$ M$_{\odot}$ (see Figure \ref{ima_MBB_NGC147}). This upper dust mass limit assumes that a possible dust reservoir has a point source geometry distributed within a single SPIRE\,250\,$\mu$m beam. The uncertainties on the upper dust mass limit are derived by varying the assumed dust temperature within the limits of uncertainties for the dust masses derived for NGC\,185 ($T_{\text{d}}$ $\sim$ 17-28\,K, see Table \ref{bestfitparameters}).
Our \textit{Herschel} observations are able to put stringent constraints on the dust reservoir in NGC\,147, improving by more than a factor of 3 in sensitivity compared to previous upper mass limits derived from MIPS\,160\,$\mu$m ($M_{\text{d}}$ $\lesssim$ 4.5 $\times$ 10$^{2}$ M$_{\odot}$, \citealt{2010ApJ...713..992M}) and ISO\,170\,$\mu$m ($M_{\text{d}}$ $\lesssim$ 4.1 $\times$ 10$^{2}$ M$_{\odot}$, \citealt{2004ApJS..151..237T}) observations for an assumed dust temperature $T_{\text{d}}$ = 20 K. The latter upper dust mass limit translates into a dust-to-stellar mass ratio $\log$ $M_{\text{d}}$/$M_{\star}$ $\lesssim$ -5.6, based on a stellar mass estimate ($M_{\star}$ $\sim$ 4.8 $\times$ 10$^{7}$ M$_{\odot}$) derived from IRAC\,3.6 and 4.5\,$\mu$m flux densities following \citet{2012AJ....143..139E}. The upper limit for the dust-to-stellar mass fraction in NGC\,147 is similar to the strongest upper limits observed for nearby galaxy samples \citep{2012A&A...540A..52C,2013MNRAS.436.1057D,2015A&A...582A.121R}.

\subsection{NGC\,185}
\begin{figure*}
\centering
\includegraphics[width=17.75cm]{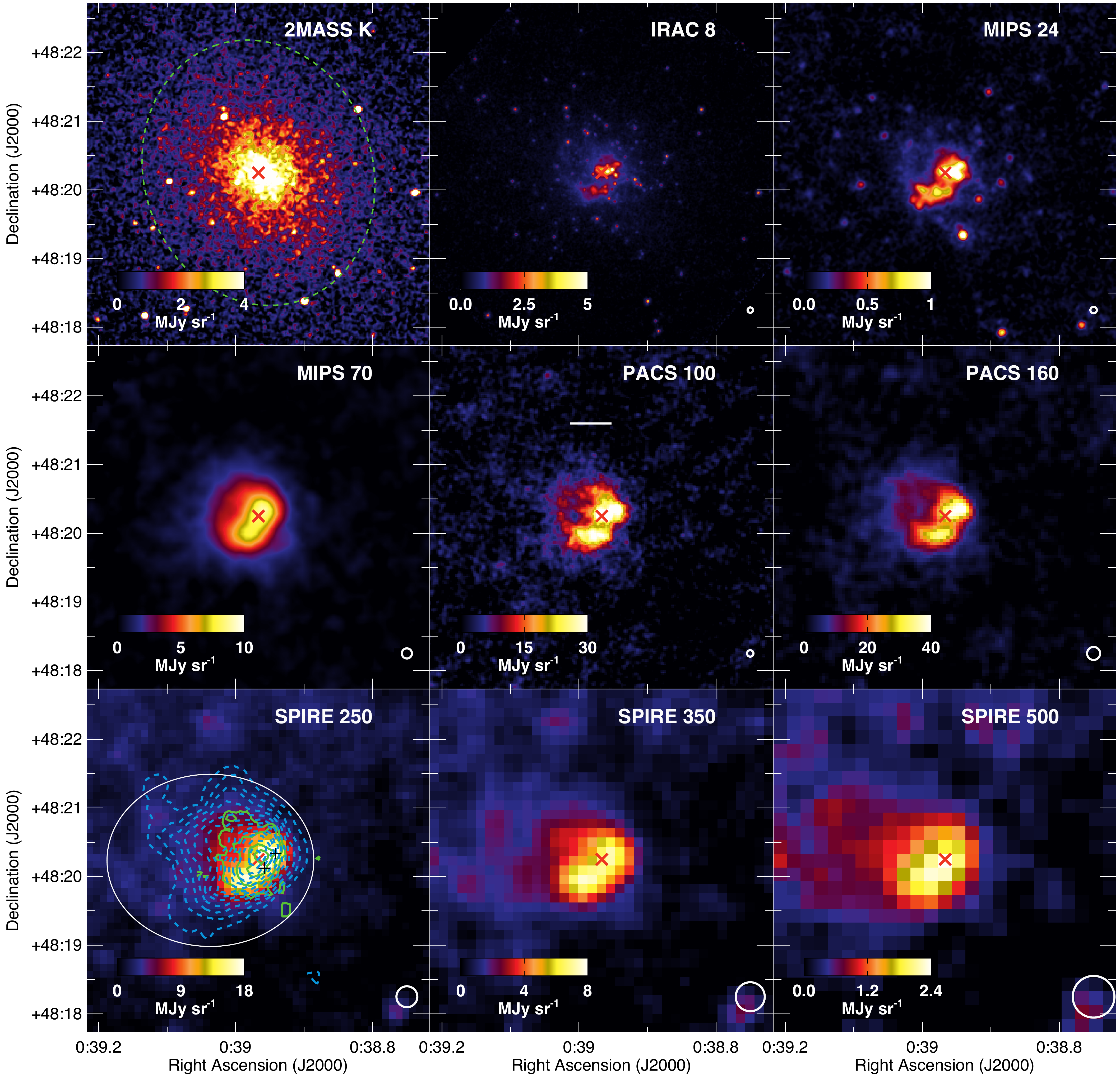}   \\
 \caption{Overview of the \textit{Spitzer} and \textit{Herschel} maps for NGC\,185. The top left panel shows a near-infrared 2MASS $K$ band image of NGC\,185 for comparison, with the ellipse indicating the extent of one-third of $B$ band 25 mag arcsec$^{-2}$ isophotal contours (i.e., $R_{\text{25}}/3$). The other panels, from left to right and top to bottom, show the dust emission in the IRAC\,8\,$\mu$m, MIPS\,24\,$\mu$m, MIPS\,70\,$\mu$m, PACS\,100\,$\mu$m, PACS\,160\,$\mu$m and SPIRE\,250, 350 and 500\,$\mu$m wavebands, respectively. The centre of the galaxy is indicated with a red cross. The FWHM of the \textit{Spitzer} and \textit{Herschel} PSFs is indicated as a white circle in the lower right corner of each panel. The white line in the PACS\,100\,$\mu$m image indicates a physical scale of 108 pc, i.e., the resolution in the SPIRE\,500\,$\mu$m map. On the SPIRE\,250\,$\mu$m map, the contours of H{\sc{i}} (FWHM $\sim$ 28$\arcsec$) and [C{\sc{ii}}] (FWHM $\sim$ 11.5$\arcsec$) observations are overlaid at their original resolution as blue dashed and green solid lines, respectively.}
              \label{Images_Herschel_NGC185}
\end{figure*}

\begin{figure*}
\centering
\includegraphics[width=18.5cm]{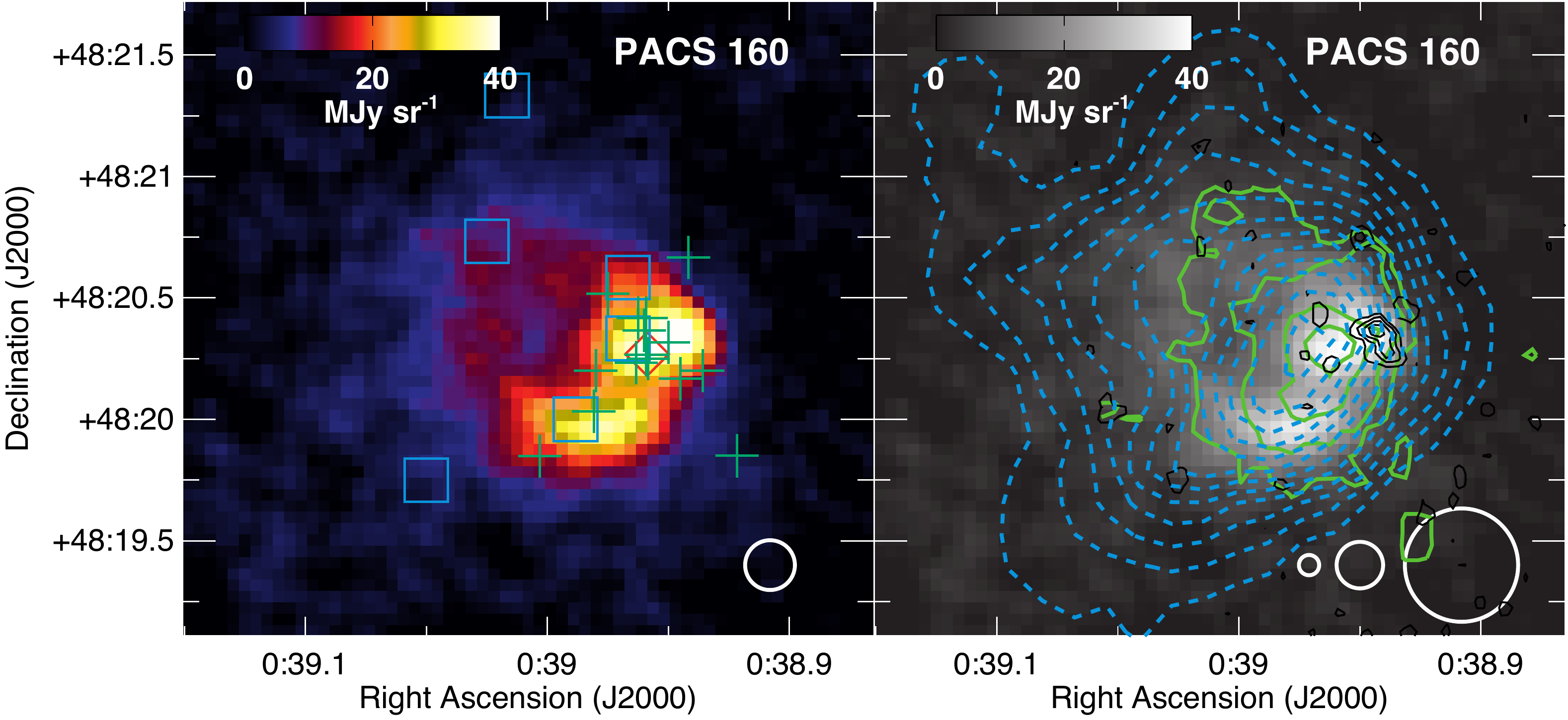}   \\
 \caption{\textit{Herschel} PACS\,160\,$\mu$m map of NGC\,185 with the size of the PACS\,160\,$\mu$m beam (FWHM $\sim$ 12.1$\arcsec$) indicated in the lower right corner of the left panel. The right panel shows, from left to right, the beam sizes of the CO (FWHM $\sim$ 5$\arcsec$), [C{\sc{ii}}] (FWHM $\sim$ 11.5$\arcsec$) and H{\sc{i}} (FWHM $\sim$ 28$\arcsec$) observations. On the left, the position of SNR-1 is indicated with a red diamond. The cyan squares represent the location of clumps in the H{\sc{i}} distribution, while the green crosses indicate the bright blue stars or star clusters first identified by \citet{1951POMic..10....7B}. In the right panel, the contours of H{\sc{i}}, CO and [C{\sc{ii}}] observations are overlaid as blue dashed, black and green solid lines, respectively, at their original resolution. All images have the same field of view (FOV).} 
               \label{Ima_comb}
\end{figure*}

Figure \ref{Images_Herschel_NGC185} shows the 2MASS $K$ band, \textit{Spitzer} (IRAC\,8\,$\mu$m, MIPS\,24\,$\mu$m, MIPS\,70\,$\mu$m) and \textit{Herschel} (PACS\,100, 160\,$\mu$m, and SPIRE\,250, 350, 500\,$\mu$m) photometry maps for NGC\,185. We achieve 1$\sigma$ noise levels of 7.7 and 9 mJy beam$^{-1}$ (equivalent to 3.7 and 2.6 MJy sr$^{-1}$) on average in the PACS\,100 and 160\,$\mu$m maps, while 1$\sigma$ rms sensitivities of 10, 11.1 and 9 mJy beam$^{-1}$ (equivalent to 0.91, 0.57 and 0.21 MJy sr$^{-1}$) are obtained in the SPIRE\,250, 350, and 500\,$\mu$m maps.
Similar to the observed morphologies at shorter (mid-)infrared wavelengths, we observe a peak in the infrared/submm emission near the centre of NGC\,185 (indicated with a red cross) in the \textit{Herschel} maps. In the PACS maps, there is an additional emission peak in the south-eastern part of the galaxy, which coincides with a shell-like morphology observed in the IRAC\,8\,$\mu$m and MIPS\,24\,$\mu$m maps \citep{2010ApJ...713..992M}.
Moving to the submillimeter wavebands covered by SPIRE, the two peaks can still be distinguished in SPIRE 250 and 350\,$\mu$m wavebands, but blend into one central peak at the resolution of the SPIRE\,500\,$\mu$m waveband. We, furthermore, observe an extended tail of diffuse dust emission on the east side of the galaxy at longer wavelengths, which is detected at a signal-to-noise (S/N) level higher than 3 (see Figure \ref{Images_Herschel_NGC185}). The extent of the diffuse dust emission partially overlaps with the broader H{\sc{i}} component observed towards the East of the galaxy (see bottom left panel in Figure \ref{Images_Herschel_NGC185} and right panel in Figure \ref{Ima_comb}), which might be an indication for a close association between the diffuse dust and gas components in NGC\,185.
The more diffuse dust emission regions towards the east detected from the \textit{Herschel} SPIRE observations were not identified from optical data. Given that the young stars are located near the centre of the galaxy and blue light is more easily susceptible to extinction processes, the reddening caused by dust clouds can more easily be identified from the central regions. Alternatively, the diffuse dust patch might correspond to foreground Galactic cirrus emission given the low galactic latitude (-14.5$^{\circ}$) of NGC\,185.

The large scale morphology of the dust component in NGC\,185 is very asymmetric, which is in a sense similar to the disturbed distribution of the more extended H{\sc{i}} gas \citep{1997ApJ...476..127Y}. The origin of this peculiar morphology of the dust component is likely related to the recent star formation activity in NGC\,185. The dust indeed coincides with the location of bright blue stars\footnote{Due to their fuzzy appearance, \citet{1999AJ....118.2229M} suggested these objects are rather young star clusters or associations with a minimum age of 100 Myr.} first identified by \citet{1951POMic..10....7B} and overlaid as green crosses on the PACS\,160\,$\mu$m map in Figure \ref{Ima_comb} (left panel). The shell structure in the south is suggested to result from a shock wave following a supernova explosion near the centre of NGC\,185 \citep{2012MNRAS.419..854G}. The remnant is thought to originate from a supernova type Ia event about 10$^{5}$ yr ago \citep{1999AJ....118.2229M}. The position of SNR-1 is shown as a red diamond in Figure \ref{Ima_comb} (left panel). 
 
Figure \ref{Ima_comb} (right panel) shows the PACS\,160\,$\mu$m map with the contours of H{\sc{i}}, CO(1-0) and [C{\sc{ii}}] 158\,$\mu$m\footnote{\citet{DeLooze_paper2} describes the [C{\sc{ii}}] dataset.} observations at their native resolution overlaid as blue dashed, black solid and green solid lines, respectively. The central dust cloud in NGC\,185 is shown to partially coincide with the molecular gas phase as traced by CO, and lies adjacent to the bulk of atomic gas mass. The southern dust shell, however, does not seem to coincide with any major peaks in the emission of the probed gas tracers (H{\sc{i}}, CO, [C{\sc{ii}}]) at first sight. From the H{\sc{i}} data cube, a clumpy concentration of H{\sc{i}} gas was found near the position of the southern dust shell (see Figure \ref{Ima_comb}, left panel). Single-dish IRAM observations, furthermore, detected CO emission from that region \citep{1997ApJ...476..127Y}, but the interferometric BIMA data only found significant CO-emitting gas near the north-west of the galaxy centre \citep{2001AJ....122.1747Y}. The molecular gas in NGC\,185, thus, appears to be mostly confined to a single cloud structure of relatively small size (34 pc $\times$ 20 pc), and does not coincide with regions where most of recent star formation activity took place. Assuming that the inclination of NGC\,185 is not edge-on (i $\sim$ 50$^{\circ}$, \citealt{2006MNRAS.369.1321D}), this offset between the young star clusters and molecular clouds confirms the break down of the Kennicutt-Schmidt (K-S) law \citep{1998ApJ...498..541K} on spatial scales $<$ 100-200 pc (e.g., \citealt{2010ApJ...722.1699S}). The breakdown of the K-S law at small scales has been argued to be due to a poor sampling of the stellar initial mass function (IMF) and gas mass function, or due to a limited duration of the molecular cloud lifetime (e.g., \citealt{2010ApJ...721..383M,2010ApJ...722L.127O,2010ApJ...722.1699S}). In NGC\,185, the supernova explosion in the centre about 10$^{5}$ years ago might have cleared the molecular gaseous material associated with the young star clusters, with the swept up molecular gas being transported by the supernova shock wave to its current location. The large line width (18.3 km s$^{-1}$) of the molecular cloud (as traced by the $^{12}$CO(1-0) rotational transition), furthermore, suggests that the cloud structure is not dynamically stable, but might endure tidal disruption processes due to its central location \citep{2001AJ....122.1747Y}. The lack of any star formation activity during the last $\sim$ 100 Myr in NGC\,185 \citep{1999AJ....118.2229M} might, thus, result from turbulence which prevents the molecular gas cloud from collapsing and forming stars.

\section{Dust SED modelling in NGC\,185}
\label{SEDmodel.sec}

\subsection{Global fluxes}

\citet{2010ApJ...713..992M} determined global flux densities for NGC\,185 within an aperture with radius R = 75$\arcsec$ centred on the optical position of the galaxy. Since this region does not include the diffuse dust emission detected in the SPIRE images towards the east of the galaxy, we choose to use an elliptical aperture (with semi-major and -minor axes of 95$\arcsec$ and 70$\arcsec$) that does capture the total dust emission originating from NGC\,185 as observed from the SPIRE\,250\,$\mu$m image (see Figure \ref{Images_Herschel_NGC185}, bottom left panel). We, furthermore, make sure that the aperture does not include any infrared-bright sources unrelated to the galaxy. We perform aperture photometry with the FITS library and utility package FUNTOOLS \citep{2011ascl.soft12002M}. Table \ref{fluxtotal} summarises the global flux densities for NGC\,185 in several wavebands. 

To determine the uncertainty on the photometry data, we take different independent noise measurements into account. The first uncertainty factor arises from the calibration, which is assumed to be uncertain by 5$\%$ in both PACS\,100 and PACS\,160 wavebands \citep{2013ExA...tmp...38B}. 
Calibration uncertainties for the \textit{Herschel} SPIRE instrument are assumed to be around 4$\%$ in each band, adding in quadrature the 4$\%$ absolute calibration error from the assumed models used for Neptune (SPIRE Observers' manual) and a random uncertainty of 1.5$\%$ accounting for the repetitive measurements of Neptune \citep{2013MNRAS.433.3062B}. The calibration in the MIPS\,24 and MIPS\,70\,$\mu$m wavebands is assumed to be uncertain by 4$\%$ \citep{2007PASP..119..994E} and 10$\%$ \citep{2007PASP..119.1019G}, respectively. For extended sources, the IRAC Instrument Handbook\footnote{http://irsa.ipac.caltech.edu/data/SPITZER/docs/irac/iracinstrumenthandbook/} recommends to assume an uncertainty factor of 10$\%$ on the calibration.
Additional uncertainty factors arise from the random background noise, instrumental noise and the confusion noise in SPIRE wavebands. The instrumental uncertainties, related to the number of scans crossing a pixel, are derived from the output error map provided as output from the data reduction. The background noise is determined as the standard deviation around the median from a set of 50 random apertures with radius R = 4$\times$FWHM in the field around the galaxy. The confusion noise due to unresolved background sources in SPIRE images is derived following Eq. 3 in \citet{2012A&A...543A.161C} and using the confusion noise estimates from \citet{2010A&A...518L...5N}. The total uncertainty is obtained as the square root of the sum of all squared independent noise contributions. In the PACS images, the total uncertainties are dominated by the random background noise (5\,$\%$) and calibration uncertainty (5\,$\%$), followed by instrumental noise (2-3\,$\%$) . The noise in the SPIRE images is dominated by uncertainties in the map making algorithm (15-20\,$\%$), with smaller contributions from confusion noise (6-9\,$\%$), background (3-4\,$\%$) and calibration uncertainties (4\,$\%$).  

\begin{table}
\caption{Total flux densities of NGC\,185 as determined from aperture photometry within an elliptical aperture 95$\arcsec$ $\times$ 70$\arcsec$ with position angle PA=0$^{\circ}$, centred on the position (RA, DEC) = (0$^{h}$:39$^{m}$:02.21$^{s}$, +48$^{\circ}$:20$\arcmin$:14.17$\arcsec$).}
\label{fluxtotal}
\centering
\begin{tabular}{lccc}
\hline 
Waveband & FWHM [$\arcsec$] & $\sigma_{\text{cal}}$ [$\%$] & $F_{\nu}$ [Jy] \\
\hline 
IRAC\,3.6\,$\mu$m & 1.7 & 10 & 0.306 $\pm$ 0.041  \\
IRAC\,4.5\,$\mu$m & 1.7 & 10 & 0.199 $\pm$ 0.029 \\
IRAC\,5.8\,$\mu$m & 1.9 & 10 & 0.271 $\pm$ 0.041\\
IRAC\,8.0\,$\mu$m & 2.0 & 10 & 0.171 $\pm$ 0.045 \\
\hline \hline 
MIPS\,24\,$\mu$m & 6 & 4 & 0.040 $\pm$ 0.016 \\
MIPS\,70\,$\mu$m & 18 & 10 & 0.697 $\pm$ 0.159  \\
PACS\,100\,$\mu$m & 6.9 & 5 &1.936 $\pm$ 0.156 \\
PACS\,160\,$\mu$m & 12.1 & 5 & 2.358 $\pm$ 0.151 \\
SPIRE\,250\,$\mu$m & 18.2 & 4 & 1.980 $\pm$ 0.322 \\
SPIRE\,350\,$\mu$m & 24.9 & 4 & 1.011 $\pm$ 0.174 \\
SPIRE\,500\,$\mu$m & 36.3 & 4 & 0.404 $\pm$ 0.089 \\ 
\hline 
\end{tabular}
\end{table}

\subsection{Flux comparison}
To verify whether the reprocessed \textit{Spitzer} maps are in agreement with the dataset presented by \citet{2010ApJ...713..992M}, we determine the flux densities within the northern, southern, and total apertures specified in Table 4 from \citet{2010ApJ...713..992M}. 
Table \ref{fluxtest} presents the flux measurements reported by \citet{2010ApJ...713..992M} (first columns) and the fluxes obtained from our reprocessed MIPS data (second columns). We find an excellent agreement between the literature fluxes and aperture photometry results determined from the reprocessed MIPS data within the error bars.

We, furthermore, compare the global photometry obtained from \textit{Herschel} observations with ancillary data at similar wavelengths.
The PACS\,100\,$\mu$m flux density (1.936 $\pm$ 0.156 Jy) corresponds well with the IRAS\,100\,$\mu$m flux measurement (1.93 $\pm$ 0.193 Jy) reported by \citet{1988ApJS...68...91R}, but is significantly higher compared to the ISO\,90\,$\mu$m flux density (0.92 $\pm$ 0.03 Jy) given by \citet{2004ApJS..151..237T}. Part of the discrepancy might be due to resolution issues and the smaller flux extraction region within the 75$\arcsec$ ISO beam. The flux inconsistency might, furthermore, be attributed to the different shapes of the bandpass filters for PACS, MIPS, IRAS and ISO. Given that accounting for the real shape of the spectrum through appropriate colour corrections would only change the fluxes by at most 10$\%$, we argue that the flux discrepancies are not driven by bandpass variations.
The PACS\,160\,$\mu$m flux density is also higher with respect to the MIPS\,160\,$\mu$m measurements (1.846 $\pm$ 0.369 Jy, \citealt{2010ApJ...713..992M}), which can be attributed to the larger aperture used for flux extraction in the \textit{Herschel} maps. Within the same area, the emission from the 160\,$\mu$m filters of MIPS and PACS is in excellent agreement (see Table \ref{fluxtest}). The ISO fluxes at 170\,$\mu$m (1.61 $\pm$ 0.21 Jy) and 200\,$\mu$m (1.01 $\pm$ 0.12 Jy) from \citet{2004ApJS..151..237T} are again lower with respect to the PACS\,160\,$\mu$m and SPIRE\,250\,$\mu$m fluxes. The limited beam size for ISO ($\sim$ 75$\arcsec$) and a possible lack of sensitivity for extended infrared emission in ISO observations are considered the main cause for these differences. 

\begin{table*}
\caption{Comparison between the aperture photometry results for MIPS 24, 70 and 160\,$\mu$m wavebands, reported by \citet{2010ApJ...713..992M} (M2010) and determined from the reprocessed MIPS data in this paper (DL2016). The last column provides the PACS \,160\,$\mu$m flux densities within the same regions.}
\label{fluxtest}
\centering
\begin{tabular}{|l|cc|cc|ccc|}
\hline 
Region & $F_{24}$ [mJy] & $F_{24}$ [mJy] & $F_{70}$ [mJy] & $F_{70}$ [mJy] & $F_{160}$ [mJy] & $F_{160}$ [mJy] & $F_{160}$ [mJy] \\
 & MIPS (M2010) & MIPS (DL2016) & MIPS (M2010) & MIPS (DL2016) & MIPS (M2010) &  MIPS (DL2016) & PACS (DL2016)  \\
\hline 
North (0.17\,arcmin$^{2}$) & 9 $\pm$ 1 & 10 $\pm$ 1 & 101 $\pm$ 20 & 108 $\pm$ 11  & 194 $\pm$ 39 & 187 $\pm$ 24 & 222 $\pm$ 35 \\ 
South (0.17\,arcmin$^{2}$) & 6 $\pm$ 1 & 7 $\pm$ 1 & 84 $\pm$ 17 & 85 $\pm$ 9 & 134 $\pm$ 27 & 134 $\pm$ 18 & 178 $\pm$ 34 \\
Total (4.91\,arcmin$^{2}$) & 46 $\pm$ 5 & 46 $\pm$ 6 & 614 $\pm$ 123 & 677 $\pm$ 86 & 1846 $\pm$ 369 & 1981 $\pm$ 285 & 2102 $\pm$ 140 \\
\hline 
\end{tabular}
\end{table*}

\subsection{Dust models}
\label{DustSED.sec}
Since the characterisation of the dust mass depends critically on the model assumptions, we apply four different dust SED models to fit the infrared/sub-millimeter SED of NGC\,185.
First of all, we perform a single MBB fit with (fixed) $\beta$ = 2.0 and dust mass absorption coefficient $\kappa_{\text{350}}$ = 0.192 m$^{2}$ kg$^{-1}$ \citep{2003ARA&A..41..241D}. We, furthermore, perform a single MBB fit with variable $\beta$. The best fitting parameters ($M_{\text{d}}$, $T_{\text{d}}$, $\beta$)\footnote{The dust mass is not strictly a free parameter, since it can be computed analytically by scaling the spectrum for the best fitting temperature and spectrum to fit the observations.} are determined through a $\chi^{2}$ minimisation using a simple gradient search method. To account for the spectral shape of the emitted spectrum after convolution with the \textit{Herschel} filters, we apply colour corrections\footnote{PACS and SPIRE colour corrections were taken from the documents http://herschel.esac.esa.int/twiki/pub/Public/PacsCalibrationWeb/cc$\_$report$\_$v1.pdf and http://herschel.esac.esa.int/Docs/SPIRE/html/spire$\_$om.html, respectively.} to the PACS\,100\,$\mu$m (1.036) and 160\,$\mu$m (0.963), and SPIRE\,250\,$\mu$m (1.0064), 350\,$\mu$m (0.9933) and 500\,$\mu$m (1.0070) fluxes, assuming an extended source and a modified blackbody spectrum with temperature 20K and spectral index $\beta$ =1.5 (which is closest to the best fitting $\beta$ value of 1.2 derived for the observed SED for NGC\,185, see Table \ref{bestfitparameters}). 

Other than the single MBB fits, we compare the observed flux densities to the emission from a full dust model composed of a mixture of grains with a specific size distribution, optical properties and grain emissivities. More specifically, we apply the \texttt{DustEm} software \citep{2011A&A...525A.103C} to predict the emission of dust grains exposed to a certain interstellar radiation field (ISRF).

\texttt{DustEm} calculates the local dust emissivity assuming non-local thermal equilibrium  (NLTE) conditions and, thus, explicitly computes the temperature distribution for every grain type of particular size and composition. The shape of the ISRF is parametrised through the ISRF observed in the solar neighbourhood \citep{1983A&A...128..212M}. 
We consider two different grain compositions in this paper. The first dust mixture corresponds to the \citet{2007ApJ...657..810D} dust model composed of polycyclic aromatic hydrocarbons (PAHs), amorphous silicates and graphites. The second dust mixture is described in \citet{2011A&A...525A.103C} and consists of PAHs, amorphous silicates and aliphatic-rich amorphous hydrocarbons (a-C:H). Both models are consistent with the observed dust SED and extinction in the diffuse interstellar medium at high-galactic latitude \citep{2011A&A...525A.103C}. The main difference between the two dust compositions is the emissivity of the carbonaceous grains in the submillimeter domain. The graphite grains in the \citet{2007ApJ...657..810D} model have a dust emissivity index $\beta \sim 1.8$ in submillimeter wavebands while the amorphous grains are characterised by a lower index of $\beta \sim 1.55$. Therefore, the same observed SED can usually be reproduced with a lower dust mass for grain mixtures including amorphous carbons compared to graphite grains \citep{2004ApJS..152..211Z,2011A&A...536A..88G,2015A&A...582A.121R}.
Fixing the shape of the ISRF and the dust grain composition, the SED fitting procedure has two free parameters: the dust mass, $M_{\text{d}}$, and the scaling factor of the ISRF, $G_{0}$ (in units of the \citealt{1968BAN....19..421H} ISRF integrated between 6 $<$ h$\nu$ $<$ 13.6 eV or 1.6 $\times$ $10^{-3}$ erg s$^{-1}$ cm$^{-2}$). We explore a parameter grid in $G_{0}$ $\in$ [0.5,10] with stepsize of 0.1 and $M_{\text{d}}$ $\in$ [1,10] $\times$ 10$^{3}$ M$_{\odot}$ increased stepwise by a factor of 1.05. We construct a pre-calculated library of dust SEDs with \texttt{DustEm} exploring the range of parameter values. Based on this library, we determine the best fitting dust SED model through a least-square fitting routine. For the $\chi^{2}$ minimisation, the model SED flux densities have been convolved with the response curves of the appropriate bandpass filters.

The main difference between the MBB fit and the full \citet{2007ApJ...657..810D} dust model is the fixed dust mass absorption coefficient in the former method, whereas the full dust model accounts for the variation in dust emissivity index and dust mass absorption coefficient at every wavelength depending on the mixture of grains with a given size distribution. For all models, the uncertainty on the best-fitting parameters is determined from a Monte Carlo method. More specifically, the same fitting routine is applied to another 1000 datasets with flux densities randomly drawn from a Gaussian distribution with the observed flux density and uncertainty as the peak and width of the Gaussian function. The 1$\sigma$ model uncertainties correspond to the 16$\%$ and 84$\%$ percentiles of the distribution of resulting parameter values. The latter parameter uncertainties rely on the assumption that the distribution of output parameters for 1000 runs with random variation in the constraining flux densities is well represented by a Gaussian function. 

\subsection{Total dust emission}
\label{GlobalSED.sec}
Before focusing on spatial variations in the strength of the ISRF and the concentration of dust mass, we model the total dust emission in NGC\,185 with global flux densities summarized in Table \ref{fluxtotal}. Because our main goal is quantifying the total dust content in NGC\,185 (rather than measuring variations in the individual dust grain types), we restrict the SED fitting procedure with the full dust model to a wavelength range 24\,$\mu$m $\leq$ $\lambda$ $\leq$ 500\,$\mu$m, which includes seven observational constraints  (MIPS\,24 and 70\,$\mu$m, PACS\,100 and 160\,$\mu$m and SPIRE\,250, 350 and 500\,$\mu$m). We do not include the IRAC\,8\,$\mu$m data point in our fitting procedure, to avoid the need for an additional free parameter (PAH fraction) in our fitting procedure. We do perform a quality check between the observed IRAC\,8\,$\mu$m flux and the model emission at 8\,$\mu$m, for a model with Galactic PAH abundance. For the modified black-body SED fitting, the wavelength range is restricted from 100 to 500\,$\mu$m to avoid the contribution of an additional warmer dust component and/or transiently heated grains at wavelengths $\leq$ 70\,$\mu$m.   

\begin{figure*}
\centering
\includegraphics[width=16.5cm]{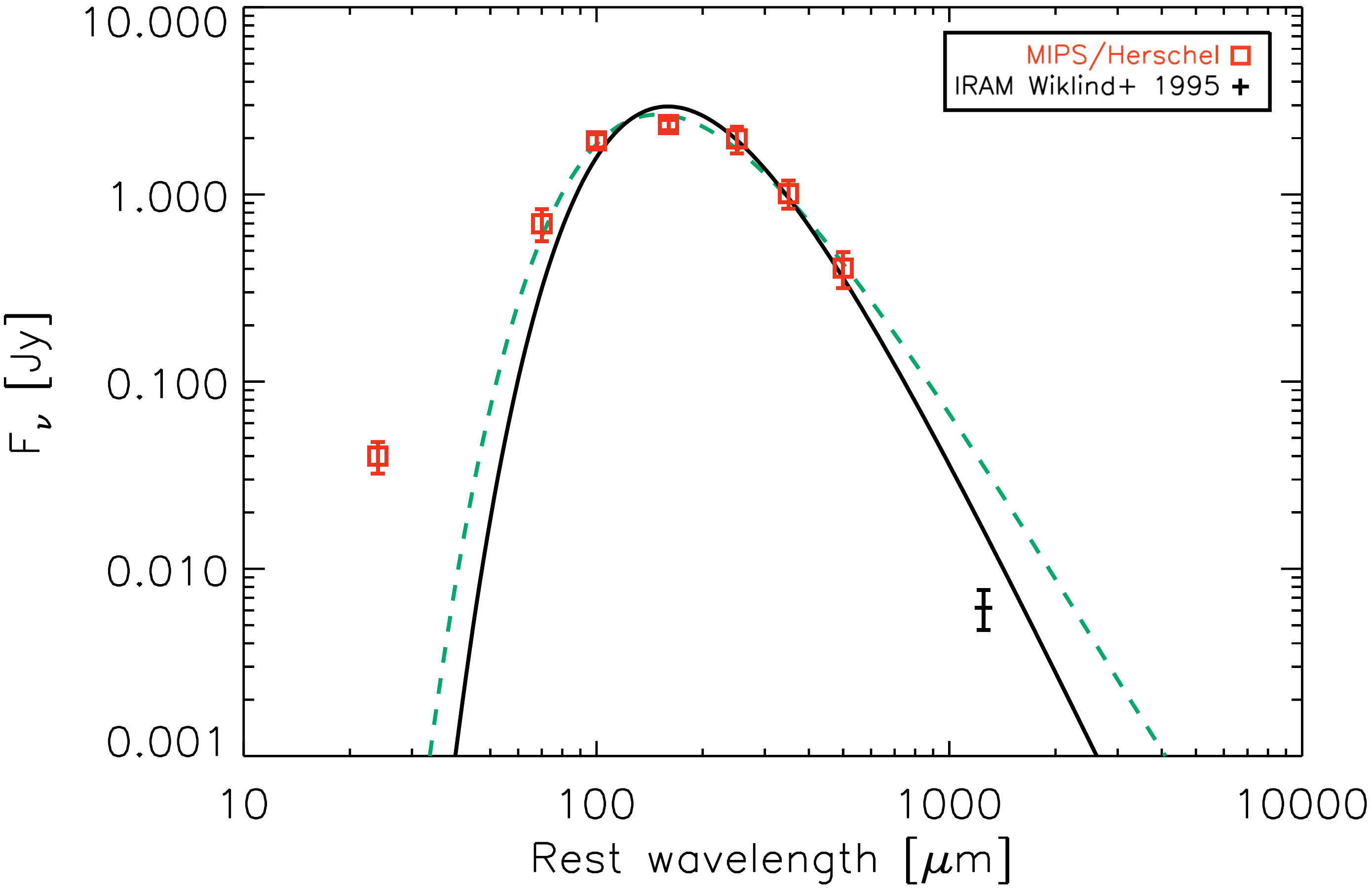}   \\
 \caption{The SED for the best-fitting modified blackbody models with fixed $\beta$=2 (black solid line) and variable $\beta$ (green dashed line). The \textit{Spitzer} MIPS and \textit{Herschel} PACS+SPIRE flux densities derived in this paper (see Table \ref{fluxtotal}), and IRAM\,1.2\,mm flux retrieved from the literature have been overlaid on the plot.}
              \label{ima_totalMBB}
\end{figure*}

\begin{figure*}
\centering
\includegraphics[width=16.5cm]{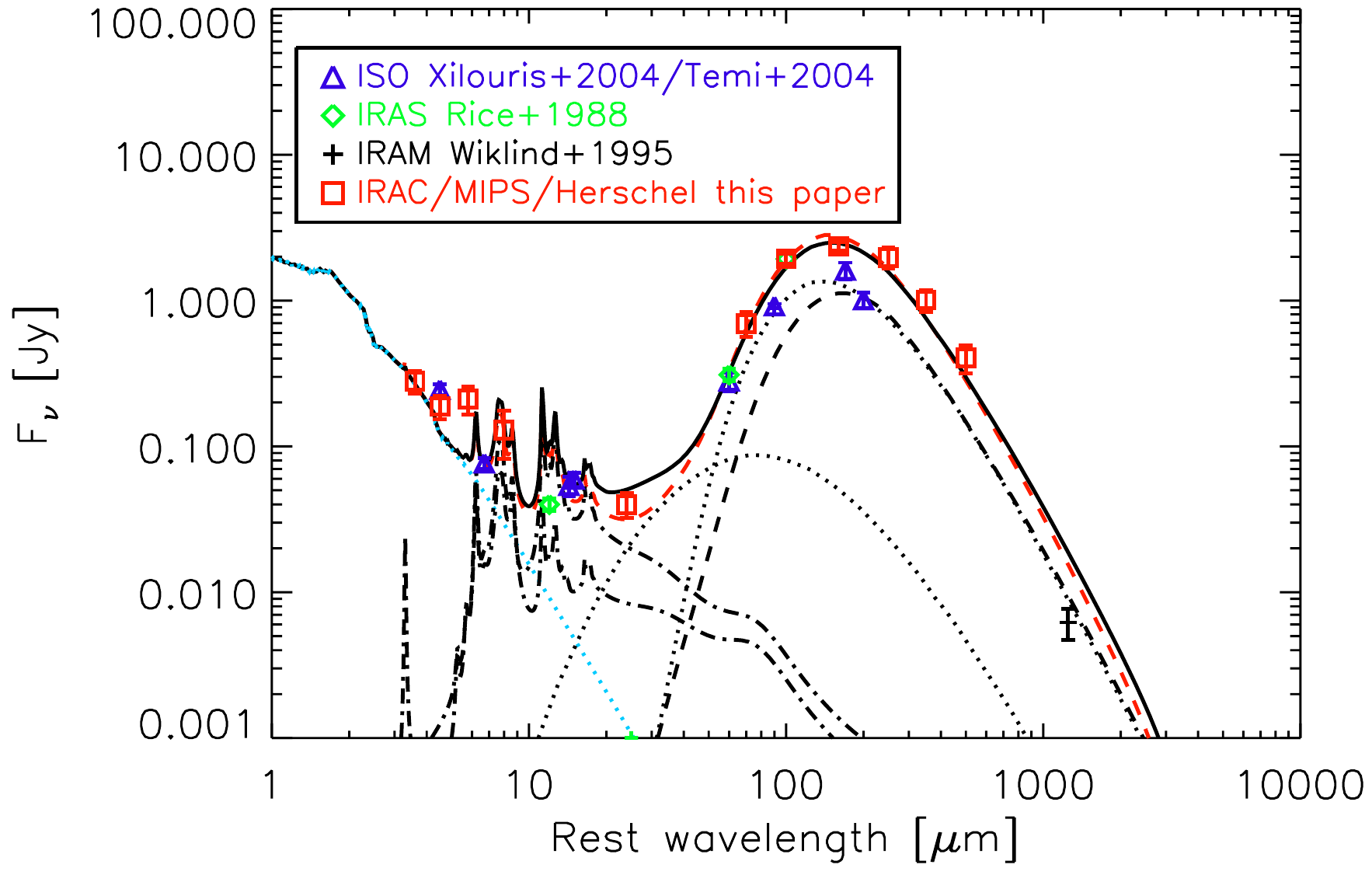}   \\
 \caption{The SED for the best-fitting \texttt{DustEm} model with a PAH+amorphous carbon+silicate (black, solid line) and PAH+graphite+silicate (red, dashed line) dust composition, fitted to measured total flux densities for NGC\,185 in the MIPS\,24, 70\,$\mu$m, PACS 100, 160\,$\mu$m and SPIRE 250, 350, 500\,$\mu$m wavebands. The legend explains the symbols used for data points from different instruments, including also the references to the relevant literature works. Individual grain species for the former dust model are indicated: neutral/ionised PAHs, small and large amorphous carbon grains, and silicate grains are represented as dotted-dashed, dotted lines, and dashed lines, respectively. A stellar component for NGC\,185 (blue, dotted curve), parametrised as a \citet{1998MNRAS.300..872M,2005MNRAS.362..799M} single stellar population with an age of 10 Gyr and a metallicity of Z = 0.005 was included in the global SED, to allow a comparison of our SED model to observations at NIR/MIR wavelengths.}
              \label{ima_totalsed}
\end{figure*}

Figure \ref{ima_totalMBB} shows the best fitting MBB models with fixed $\beta$=2 (black solid line) and variable $\beta$ (green dashed line).
Figure \ref{ima_totalsed} shows the best fitting SED model derived for dust models with a PAH+amorphous carbon+silicate (black solid line) and PAH+graphite+silicate (red dashed line) composition. The total flux densities measured from different observations for NGC\,185 are overlaid with different symbols, and explained in the legends of Figures \ref{ima_totalMBB} and \ref{ima_totalsed}. 
The best fitting parameters for the different SED fits are summarized in Table \ref{bestfitparameters}. For the full SED models, the dust temperature estimate is obtained from the ISRF scaling factor $G_{\text{0}}$ through $T_{\text{d}}$ = 19.7 K $\times$ $G_{\text{0}}^{1/4+\beta}$ with $\beta$ = 2 and the average dust temperature in the Milky Way galaxy (19.7 K, \citealt{2014Planck}). 

The output dust mass and temperature from the full SED dust models and the modified blackbody fit with variable $\beta$ are consistent within the error bars. The modified blackbody SED fit with fixed $\beta$ is an exception with a higher dust mass content and a lower dust mass temperature. The over- and under-prediction of the PACS\,160\,$\mu$m and 100\,$\mu$m emission by 25$\%$ and 19$\%$, respectively, in the fixed $\beta$ MBB model implies that the higher dust mass and lower dust temperature are not realistic for the dust population in NGC\,185. The low emissivity index $\beta$=1.2 that seems favourable to fit the width and Rayleigh-Jeans slope of the dust SED in NGC\,185 is consistent with the value $\beta$ of 1.2 derived from ISO observations for NGC\,185 by \citet{2004ApJS..151..237T}. The low $\beta$ value fitted to the SED of NGC\,185 is also consistent with the tendency for lower $\beta$ values observed in low metallicity objects (e.g., \citealt{2012A&A...540A..54B,2013A&A...557A..95R}). \citet{2015A&A...582A.121R} also finds a trend of broader dust SED towards low metallicity in a local galaxy sample, which they attribute to a wider range of dust equilibrium temperatures due to a clumpy ISM structure. The PACS\,70\,$\mu$m flux density is also consistent with the MBB model with $\beta$ = 1.2 and $T_{\text{d}}$ $\sim$ 23.5\,K, which suggests that the SED model captures the distribution of different dust temperatures better than for the MBB model with $\beta$ = 2 (see Figure \ref{ima_totalMBB}). The dust mass derived from the MBB fit with variable $\beta$ (5.0$^{+1.7}_{-1.3}$ $\times$ 10$^3$ M$_{\odot}$) is also consistent with the dust mass derived from the full SED models (5.0$\pm$1.0 $\times$ 10$^3$ M$_{\odot}$). We, however, need to compare these dust mass derivations with caution since the spectral index of the best fitting SED model ($\beta$ = 1.2) and the dust model (with $\beta$ = 1.8) used to characterise the normalisation factor of the SED fit ($\kappa_{350}$ = 0.192 m$^{2}$ kg$^{-1}$) are not compatible, what might affect the dust mass determination \citep{2013A&A...552A..89B}.

Comparing the full SED models with a \citet{2007ApJ...657..810D} or \citet{2011A&A...525A.103C} dust composition, the scaling of the radiation field is higher by 40$\%$ for the \citet{2007ApJ...657..810D} dust model compared to the \citet{2011A&A...525A.103C} dust composition. The warmer radiation field derived for the \citet{2007ApJ...657..810D} is thought to result from the lower dust emissivity in the mid-infrared wavelength domain for the same hydrogen column compared to the emissivity of grains in the \citet{2011A&A...525A.103C} model. With the MIPS\,24\,$\mu$m data point being the single constraining data point in the mid-infrared wavelength regime\footnote{We refrain from using WISE data since an unidentified source has been detected in the WISE\,22\,$\mu$m frame at a distance of 65$\arcsec$ (or 200 pc) from the optical centre of NGC\,185, which was not earlier identified in MIPS\,24\,$\mu$m maps. Blending issues prevent an accurate measurement of the WISE\,22\,$\mu$m flux density for NGC\,185 after convolution of the maps.}, it will have a strong influence on the scaling factor of the radiation field. Since a whole range of dust temperatures can be expected along a single sight line, we do not interpret this difference in radiation field (or thus dust temperature) as a significant difference between the two models. 

The dust mass derived from the global SED fitting procedures combining \textit{Spitzer} and \textit{Herschel} data is more than a factor of 2-3 higher compared to previous dust mass estimates based on ISO (1.5 $\times$ 10$^3$ M$_{\odot}$ for a dust temperature $T_{\text{d}}$ = 22 K, \citealt{2004ApJS..151..237T}) and \textit{Spitzer} data (1.9$^{+1.9}_{-0.9}$ $\times$ 10$^3$ M$_{\odot}$ for a dust temperature $T_{\text{d}}$ = 17 K, \citealt{2010ApJ...713..992M}), which confirms the need for dust emission constraints longward of 200\,$\mu$m to recover the cold dust component in galaxies \citep{2010A&A...518L..65B,2012MNRAS.419.1833B,2010A&A...518L..89G,2011A&A...532A..56G,2012MNRAS.423.2359D,2015A&A...582A.121R}. 
The detection of extended dust emission in the SPIRE wavebands indeed suggests that a more diffuse, low temperature dust component is present in NGC\,185, which was not identified by previous far-infrared observations. The longer wavelength coverage is, however, not the only reason for the difference in dust mass. Also the different model assumptions and fitting procedure play a role. The dust mass estimates of \citet{2004ApJS..151..237T} were derived from a single modified black-body fit (MBB), while \citet{2010ApJ...713..992M} fitted the SED with the BARE-GR-S model of \citet{2004ApJS..152..211Z} (including bare graphite grains, bare silicate grains, and PAHs) with the dust mass and the intensity of the radiation field as free parameters. The dust temperature (21.4 K) derived from the MBB fit to ISO data points is higher than the dust temperatures derived from the \textit{Spitzer}+\textit{Herschel} data points, which explains the lower dust masses reported by \citet{2004ApJS..151..237T} and confirms the need for longer wavelength data to trace the colder dust. To compare our model results with \citet{2010ApJ...713..992M}, who do not have submillimeter constraints, we use the \citet{2011A&A...525A.103C} dust mixture to model the MIPS (24, 70, 160) $\mu$m global flux measurements reported by \citet{2010ApJ...713..992M}. We derive a total dust mass of $M_{\text{d}}$ = 2.5$^{+1.2}_{-0.3}$ $\times$ 10$^{3}$ M$_{\odot}$, which is compatible with the dust mass (1.9$^{+1.9}_{-0.9}$ $\times$ 10$^3$ M$_{\odot}$) derived by \citet{2010ApJ...713..992M}, and confirms that longer wavelength sub-millimeter data are necessary to accurately trace the cold dust component in galaxies. 

The average dust temperature in NGC\,185 ($T_{\text{d}}$ $\sim$ 21-22 K) is in agreement with the range of dust temperatures derived for nearby late-type galaxies (e.g., \citealt{2012ApJ...745...95D,2014MNRAS.440..942C,2014A&A...565A.128C}). The dust temperature is, however, significantly lower than average grain temperatures ($T_{\text{d}}$ $\sim$ 32 K) derived for a sample of low-metallicity star-forming dwarf galaxies (e.g., \citealt{2013A&A...557A..95R}) and also on the low edge of the range of dust temperatures derived for a sample of more massive early-type galaxies from the \textit{Herschel} Reference Survey (23.9$\pm$0.8 K, \citealt{2012ApJ...748..123S}). Given the low level of star formation activity in NGC\,185 \citep{1999AJ....118.2229M}, these dust temperatures are a natural consequence of the low energy input from young stars.

Although the PAH abundance was fixed to the Galactic value in the full SED models, we can check how the fraction of polycyclic aromatic hydrocarbons in NGC\,185 compares with the Galactic average ($f_{\text{PAH}}$ $\sim$ 4.5$\%$). Comparing the IRAC\,8\,$\mu$m emission from observations (0.13 $\pm$ 0.05 Jy) with the dust SED model (0.10 Jy), we do not find any evidence for a reduced PAH abundance in NGC\,185, which has often been observed in low-metallicity galaxies (e.g., \citealt{2004A&A...428..409B,2005ApJ...628L..29E,2006ApJ...646..192J,2006A&A...446..877M,2007ApJ...663..866D,2008ApJ...678..804E,2008ApJ...672..214G}). The critical metallicity for the drop in PAH abundance has, however, been suggested to be around 12+$\log$(O/H) $\sim$ 8.1-8.2, which might explain why we do not see a similar effect in NGC\,185 with 12+$\log$(O/H) $\sim$ 8.25. The weak radiation field compared to metal-poor star-forming dwarfs might, furthermore, not be sufficient to destroy PAHs in NGC\,185. 
Also the high supernova activity (based on the high [Fe {\sc{ii}}]/[Ne {\sc{ii}}] $\sim$ 1-2, \citealt{2010ApJ...713..992M}) and the delayed injection of dust from AGB stars compared to supernova \citep{2012MNRAS.419..854G} does not seem to cause PAH destruction by supernova shocks (e.g., \citealt{2006ApJ...641..795O}) or a delayed injection of carbon rich-dust (e.g., \citealt{2008ApJ...672..214G}). The detection of several PAH features in the IRS spectra \citep{2010ApJ...713..992M} supports the presence of PAH molecules in NGC\,185. The latter result confirms previous statements that the PAH abundance is more sensitive to the ionisation parameter rather than the metal abundance in galaxies \citep{2008ApJ...682..336G}.  

\begin{table}
\caption{Overview of the output parameters ($\beta$, $T_{d}$, $M_{\text{d}}$) and ($G_{0}$, $T_{d}$, $M_{\text{d}}$) corresponding to the best-fitting MBB fit (top) and SED model for PAH+graphite+silicate and PAH+amorphous carbon+silicate dust compositions (bottom), respectively. The last column presents the reduced $\chi^{2}$ value that corresponds to the SED fit.}
\label{bestfitparameters}
\centering
\begin{tabular}{lcccc}
\hline 
MBB models & $\beta$ & $T_{d}$ [K] & $M_{d}$ [M$_{\odot}$] & $\chi^{2}$ \\
\hline
Fixed $\beta$  & (2.0) & 18.2$^{+0.6}_{-0.6}$ & 9.0$^{+1.6}_{-1.4}$ $\times$ $10^{3}$ & 0.17 \\
Variable $\beta$ & 1.2$^{+0.4}_{-0.4}$  & 23.5$^{+3.4}_{-3.4}$ & 5.0$^{+1.7}_{-1.3}$ $\times$ $10^{3}$ & 0.11 \\
\hline 
\hline 
Full dust models & $G_{0}$ & $T_{d}$ [K] & $M_{d}$ [M$_{\odot}$] & $\chi^{2}$ \\
\hline
PAH+graphite+silicate & 2.1$^{+0.3}_{-0.3}$ & 22.3$^{+0.5}_{-0.6}$ & 5.0$^{+0.8}_{-0.7}$ $\times$ $10^{3}$ & 0.85 \\
PAH+am. carbon+silicate & 1.5$^{+0.3}_{-0.3}$ & 21.1$^{+0.6}_{-0.8}$ & 5.2$^{+0.8}_{-0.7}$ $\times$ $10^{3}$ & 0.77 \\
\hline 
\end{tabular}
\end{table}

\begin{figure}
\centering
\includegraphics[width=8.5cm]{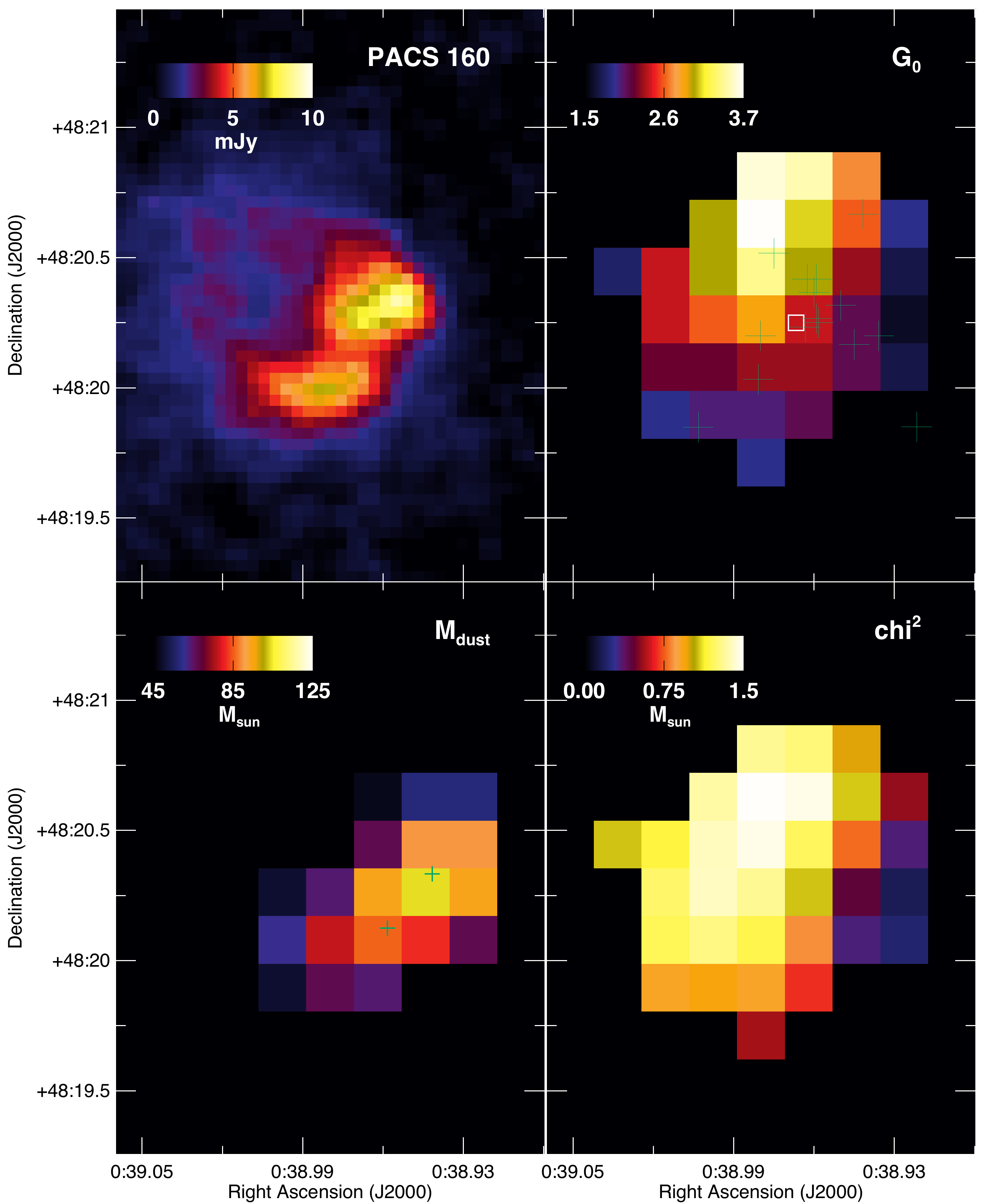}   \\
 \caption{The parameter maps for the radiation field strength $G_{0}$ (top right), dust mass $M_{d}$ (bottom left) and reduced $\chi^{2}$ (bottom right) corresponding to the best fit from a pixel-by-pixel SED fitting (with 36$\arcsec$ or 108 pc sized pixels) procedure with DustEm for a PAH+amorphous carbon+silicate dust composition. The PACS\,160\,$\mu$m map of NGC\,185 is presented on the same scale for comparison in the top left panel. The optical centre and the position of young blue stars in NGC\,185 are indicated as a white rectangle and green crosses, respectively, in the top right panel, whereas the green crosses in the bottom left panel show the location of two dust clouds identified from optical observations.}
              \label{ima_sedpar}
\end{figure}

\subsection{Sub-millimeter excess}
\label{Excess.sec}
An excess emission with respect to dust models has been identified at sub-millimeter wavelengths (sometimes even in millimeter wavebands) in several low-metallicity star-forming dwarfs (e.g., \citealt{2003A&A...407..159G,2005A&A...434..867G,2009A&A...508..645G,2011A&A...532A..56G,2010A&A...518L..52G,2012A&A...537A.113P,2013ApJ...778...51K,2013A&A...557A..95R,2014ApJ...797...85G,2015A&A...574A.126G}) and spiral galaxies (e.g., \citealt{2006ApJ...652..283B,2009ApJ...706..941Z}). The submm excess seems to preferentially occur in low-metallicity environments, but a correlation between the strength of the excess and metal abundance could not yet be identified \citep{2013ApJ...778...51K,2013A&A...557A..95R,2015A&A...582A.121R}. Also on the origin of the excess emission, several studies have not yet been able to converge to a conclusion. The excess emission has been attributed to the presence of a cold dust reservoir \citep{2003A&A...407..159G,2005A&A...434..867G,2009A&A...508..645G}, non-thermal emission of spinning dust particles in ionised media \citep{2010A&A...523A..20B}, magnetic nano-particles composed of metallic Fe or Fe oxides \citep{2012ApJ...757..103D}, variations in the spectral index with temperature \citep{2009A&A...506..745P}, an increase of the total emissivity of grains in colder, denser media due to grain coagulation \citep{2003A&A...398..551S,2009A&A...506..745P,2011A&A...536A..25P}, or an enhanced population of small hot dust grains \citep{2002A&A...382..860L,2009ApJ...706..941Z}.

For the different modelling techniques applied to the dust SED of NGC\,185 (see Section \ref{GlobalSED.sec}), the excess emission is defined as $E_{\lambda} = \frac{F_{\lambda}^{\text{obs}} - F_{\lambda}^{\text{model}}}{F_{\lambda}^{\text{model}}}$. The excess is considered significant if $F_{\lambda}^{\text{obs}}$ - $F_{\lambda}^{\text{model}}$ $>$ $\sigma_{\lambda}^{\text{obs}}$ where $F_{\lambda}^{\text{obs}}$ and $F_{\lambda}^{\text{model}}$ are the observed and modelled flux density with $\sigma_{\lambda}^{\text{obs}}$ the uncertainty on the observed flux density at the wavelength $\lambda$. At the longest \textit{Herschel} SPIRE waveband of 500\,$\mu$m, we do not detect any excess emission based on the best fitting single modified blackbody SEDs or the full dust model with a PAH+amorphous carbon+silicate dust composition. For the full dust model with a PAH+graphite+silicate dust composition, we find a marginally significant excess emission of $E_{\text{500}}$ = 34$\%$. Given that the excess in the SPIRE\,500\,$\mu$m waveband appears to be model dependent, we do not consider the excess emission as significant. The absence of a submm excess in NGC\,185 is consistent with the lack of any submm/mm excess emission in NGC\,205 \citep{2012MNRAS.423.2359D}.

The total flux density (6.2 $\pm$ 1.5 mJy) at millimeter wavebands (1250\,$\mu$m) reported by \citet{1995A&A...297L..71W} is about a factor of 2.5 lower than the extrapolation of our best fitting models to millimeter wavelengths. With the millimeter observations of \citet{1995A&A...297L..71W} covering only the inner $\sim$1\,$\arcmin$ region of NGC\,185, the observations might possibly miss millimeter emission residing from the diffuse dust clouds in the east of the galaxy detected in the SPIRE maps. Since about 62$\%$ of the total SPIRE 500\,$\mu$m emission is emitted from the central $\sim$1\,$\arcmin$ of NGC\,185 (at a resolution of $\sim$ 36.3$\arcsec$), we argue that a smaller coverage can not be the only reason for the discrepant millimeter data point. With a beam throw of 55$\arcsec$ in azimuth direction to recover the sky background, the sky subtraction might have been biased by more extended millimeter radiation from NGC\,185. Alternatively, the flux determination might have been affected by calibration uncertainties and/or inaccuracies in the image reconstruction.

\subsection{Spatially resolved modelling}
\label{ResolvedSED.sec}
Based on resolved SED modelling, we investigate possible variations in the strength of the radiation field and locate the most massive infrared dust clouds. We perform a pixel-by-pixel SED fit at the resolution of the SPIRE\,500\,$\mu$m waveband (FWHM $\sim$ 36.3$\arcsec$ or 108 pc) with pixel size 12$\arcsec$ (or 36 pc). The SED fitting procedure is again constrained by the emission from MIPS\,24\,$\mu$m and 70\,$\mu$m, PACS 100\,$\mu$m and 160\,$\mu$m, and SPIRE 250\,$\mu$m, 350\,$\mu$m, and 500\,$\mu$m wavebands. Only pixels with $>$ 2$\sigma$ detections in at least five of those bands (i.e., sufficient to cover the peak of the SED) are used in the fitting procedure. 
Figure \ref{ima_sedpar} (bottom row) shows the parameter maps for the ISRF scaling factor, $G_{0}$, and dust mass, $M_{d}$, for the dust model composed of PAH+amorphous carbon+silicate grains. The reduced $\chi^{2}$ value for the fit in every pixel is shown in the bottom right panel. The output maps derived for the PAH+graphite+silicate dust model \citep{2007ApJ...657..810D} show similar patterns in parameter variation throughout the galaxy and are, therefore, not shown here. As a comparison, the PACS\,160\,$\mu$m is shown in the top left panel of Figure \ref{ima_sedpar}.

The highest concentration of dust mass in NGC\,185 is found north-west and south-east of the centre, coinciding with the position of the emission peaks in the PACS\,160\,$\mu$m image and the two dust clouds identified from optical observations \citep{1963AJ.....68..691H,1981AJ.....86.1312G,1998JKAS...31...51K}.  It is difficult to compare the total dust mass ($M_{\text{d}}$~=~1.8 $\times$ 10$^{3}$ M$_{\odot}$) obtained from summing individual pixels with the global dust mass ($M_{\text{d}}$~=~5.1 $\times$ 10$^{3}$ M$_{\odot}$) due to the significantly smaller area covered in the spatially resolved SED fitting procedure. With the total dust mass derived from the global fitting almost three times larger compared to the dust mass obtained from the resolved fit, most of the cold dust in NGC\,185 is thought to be co-spatial with the extended emission observed at sub-millimeter wavebands. If the spatial variation in dust temperatures in NGC\,185 is not well represented by the global dust SED fit, this might result in somewhat too high dust mass estimates. We, however, believe that the extended emission in the SPIRE wavebands is detected at sufficient signal-to-noise level and belongs to NGC\,185 due to its association with the H{\sc{i}} component in the galaxy rather than it being Galactic cirrus emission. We will, therefore, use the global dust mass estimate derived from the total emission in NGC\,185 in the remainder of this paper. 

The ISRF appears to be stronger North of the centre of NGC\,185. The $G_{0}$ range covered in NGC\,185 (1.5 $\lesssim$ $G_{\text{0}}$ $\lesssim$ 3.7) corresponds approximately to a variation in dust temperatures between 21 K and 24.5 K, and is very similar to the radiation field strengths observed in NGC\,205 (0.4 $\lesssim$ $G_{\text{0}}$ $\lesssim$ 3, \citealt{2012MNRAS.423.2359D}). Since the emission of both old and young stellar populations dominates near the centre of the galaxy, and the uncertainties on the SED fits are larger in the North, we argue that the observed trend in G$_{\text{0}}$ does not have a physical interpretation but rather is driven by larger uncertainties on the SED modelling results towards the North.
  
\section{Dust production and destruction}
\label{DustSource.sec}
We make an inventory of the different sources responsible for dust injection into the ISM of NGC\,147 and NGC\,185 (see Section \ref{DustProduction}). Both supernovae and asymptotic giant branch stars are considered to dominate the injection of heavy elements into the ISM. We, furthermore, repeat the same exercise for the dwarf spheroidal galaxy, NGC\,205 with a dust mass $M_{\text{d}}$ $\sim$ 1.1-1.8 $\times$ 10$^{4}$ M$_{\odot}$ and dust temperature $T_{\text{d}}$ $\sim$ 18-22 K, as measured by \citet{2012MNRAS.423.2359D}. We also calculate the dust lifetime for each of the galaxies based on their ISM mass and supernova rate )see Section \ref{DustLifetime}). Section \ref{PosCaveats} discusses the uncertainties on the calculations of dust mass production and dust survival times. Comparing the current dust content with the dust production over a time period shorter than the typical dust survival time allows us to draw important conclusions about dust evolution processes in dwarf spheroidal galaxies (see Section \ref{DustImplications}).

\subsection{Dust lifetime}
\label{DustLifetime}
We calculate the dust lifetime for NGC\,185 and NGC\,205 based on the prescriptions from \citet{2011A&A...530A..44J}, who have re-evaluated the dust survival lifetimes based on the original recipes from \citet{1989IAUS..135..431M} and \citet{1994ApJ...433..797J,1996ApJ...469..740J}. Relying on the total ISM mass, $M_{\text{ISM}}$ (= $M_{\text{d}}$ + $M_{\text{g}}$), and the effective interval between consecutive supernova explosions, $\tau_{\text{SN}}$, the dust lifetime is calculated through (see Eq. 7 in \citealt{2011A&A...530A..44J}):
\begin{equation}
\label{EqLifetime}
t [\text{yr}] = \frac{(M_{\text{ISM}} [\text{M}_{\odot}] * \tau_{SN} [\text{yr}])}{2 * 2194 * (1.1/n)}
\end{equation} 
where $n$ = 6 for silicate/oxygen-rich dust and $n$ = 3 for carbonaceous dust. The ISM masses are calculated based on the dust masses derived from \textit{Herschel} observations for NGC\,185 ($M_{\text{d}}$ = 5.1 $\times $10$^{3} $ M$_{\odot}$, see this paper) and NGC\,205 ($M_{\text{d}}$ = 1.1-1.8 $\times$ 10$^{4}$ M$_{\odot}$, \citealt{2012MNRAS.423.2359D}), and the updated gas mass measurements for NGC\,185 ($M_{\text{g}}$ = 2-5 $\times$ 10$^{5}$ M$_{\odot}$) and NGC\,205 ($M_{\text{g}}$ = 9-29 $\times$ 10$^{5}$ M$_{\odot}$) from \citet{DeLooze_paper2}. 

For NGC\,185, we rely on the SN rates N$_{\text{SN II+Ib}}$ $\sim$ 2.3 $\times$ 10$^{-6}$ yr$^{-1}$ derived by \citet{1999AJ....118.2229M}. By assuming a similar rate of core-collapse and type Ia supernovae, \citet{1999AJ....118.2229M} find a total SN rate of N$_{\text{SN}}$ $\sim$ 4.6 $\times$ 10$^{-6}$ yr$^{-1}$ in NGC\,185\footnote{The latter estimate for the total SNe rate is four times higher compared to the results of the chemical evolution models of \citet{2012MNRAS.419.3159M}. We prefer to rely on the higher SN rate estimate from \citet{1999AJ....118.2229M} to derive an upper limit on the contribution of SNe to the dust mass production in NGC\,185. The dust lifetime in NGC\,185 could, thus, be a factor of 4 higher based on the lower SN rate from \citet{2012MNRAS.419.3159M}.}, which results in a time interval of 2.2 $\times$ 10$^{5}$ yr between supernova explosions. 
For NGC\,205, we estimate the SNe II+Ib supernova rate (N$_{\text{SN II+Ib}}$ $\sim$ 2.4 $\times$ 10$^{-6}$ yr$^{-1}$) from the SFR $\sim$ 7.0 $\times$ 10$^{-4}$ M$_{\odot}$ yr$^{-1}$ calculated based on color-magnitude diagrams by \citet{2009A&A...502L...9M}. For a similar fraction of core-collapse and type Ia explosions, we derive a total SN rate N$_{\text{SN}}$ $\sim$ 4.8 $\times$ 10$^{-6}$ yr$^{-1}$ and typical time interval of 2.1 $\times$ 10$^{5}$ yr between supernova explosions. 

Based on those ISM masses and SN rates, we derive typical dust lifetimes of 21-52\,Myr and 42-104\,Myr for carbonaceous and silicate grains in NGC\,185, respectively. The typical dust survival time for carbonaceous grains and silicates in NGC\,205 is estimated to be around 90-286\,Myr and 179-572\,Myr, respectively. In our Galaxy, dust lifetimes are estimated to be around 200\,Myr and 400\,Myr for carbonaceous and silicate grains, respectively \citep{1994ApJ...433..797J,1996ApJ...469..740J,2008A&A...492..127S,2011A&A...530A..44J}. A re-evaluation of the dust processing by supernova shocks by \citet{2014A&A...570A..32B} resulted in shorter dust lifetimes for carbon grains (62\,Myr) and silicate grains (310\,Myr). Recent simulations by \citet{2015ApJ...803....7S} for evolving, radiative supernova remnants (as opposed to the plane parallel steady shock models used in the past), however, find much longer destruction timescales of 3.2\,Gyr and 2.0\,Gyr for grains made of carbon and silicate material. The increase in dust destruction time scale by about one order of magnitude compared to previous work can be attributed to the local ISM conditions (i.e., \citealt{2015ApJ...803....7S} assumes that the supernova shocks evolve in a medium dominated by a warm ISM phase), revised estimates of supernova rates and ISM mass and the inclusion of the effects of the hydrodynamical shock evolution.

The dust lifetimes for carbon grains (21-285\,Myr) and silicates (42-569\,Myr) derived for the dwarf spheroidal galaxies in the Local Group\footnote{Due to the absence of any star formation activity during the last $\sim$ 1 Gyr \citep{1997AJ....113.1001H}, we assume that (core-collapse) supernova explosions had a negligible effect on the dust destruction in NGC\,147 during the last few hundred million years and we, therefore, can not make any predictions for the lifetime of dust residing in NGC\,147 based on Equation \ref{EqLifetime}.} are on average lower compared to those latter values (although the upper limits on the dust lifetime for carbon and silicate grains in NGC\,205 could be consistent with the MW values), indicating that the dust destruction timescales might be shorter in low-mass objects with low metal abundance.

\subsection{Dust production}
\label{DustProduction}
Based on the total dust mass injection rates derived for the Large Magellanic Clouds (\citealt{2012ApJ...753...71R}, see their Table 9), we estimate dust mass injection rates (see Table \ref{dustmassloss}) for C-rich AGBs, O-rich AGBs and red supergiants (RSG) scaled to the fraction of the observed AGB populations \citep{2005AJ....130.2087D} in the three dwarf spheroidal satellites of M\,31\footnote{Due to the lack of constraints on the O-rich AGB and RSG population, we assume a similar O-rich-to-total AGB and RGB-to-total AGB ratio as observed in the LMC.}. 
We scale the dust mass injection rates derived for the LMC with the number of C-rich AGB stars identified in the dwarfs (see Table 5 in \citealt{2005AJ....130.2087D}, adding up the C stars identified in the inner and outer galaxy regions) relative to the C-rich AGB population in the LMC identified based on the GRAMS model grid (i.e., 7281 objects, see \citealt{2012ApJ...753...71R}). Based on the 103, 387, and 65 C stars identified in NGC\,185, NGC\,205 and NGC\,147, we derive dust mass injection rates of 1.93, 7.25, 1.22 $\times$ 10$^{-7}$ M$_{\odot}$ yr$^{-1}$ for C-rich AGB stars, respectively. The dust mass loss rates for O-rich AGB and RGB stars are estimated by scaling the fractional dust mass injection rates of C-rich stars (64.6$\%$) to the fraction of dust mass injected by O-rich AGB (26$\%$) and RSG (9.4$\%$) stars in the LMC. Table \ref{dustmassloss} (top part) gives an overview of the estimated dust mass loss rates of C-rich and O-rich AGB stars, and RSG stars. The second part of Table \ref{dustmassloss} gives an overview of the dust mass that was lost by AGB stars during a period equal to the estimated dust lifetimes (see Section \ref{DustLifetime}). We assume here that C-rich AGB stars mostly produce carbon dust while O-rich AGB stars and RSGs mainly condense silicate grain material. For NGC\,147, we assume a dust lifetime of 1\,Gyr, which represents the timescale since the last star formation activity in this galaxy. 

To calculate the injection of elements from core-collapse supernovae (type SNe II + SN Ib), we rely on the observed dust mass of $\sim$0.23 M$_{\odot}$ near the centre of the remnant SN 1987A and, thus, unaffected by shock destruction \citep{2014ApJ...782L...2I}. The latter estimate for the produced dust mass in core-collapse supernovae is compatible with the dust masses (0.1-0.3 M$_{\odot}$) derived based on theoretical models \citep{2001MNRAS.325..726T} and observed in another SNe type II, the Crab Nebula \citep{2012ApJ...760...96G}. Because the dust mass produced in supernovae depends on the mass of the progenitor, and the density of the surrounding ISM will determine the efficiency of grain destruction in the reverse shock, the dust mass produced in core-collapse supernova remains very uncertain. Theoretical models suggest that only a small fraction ($\lesssim$ 0.1 M$_{\odot}$) of the original dust mass produced in core-collapse supernovae is able to survive the passage of the reverse shock \citep{2015arXiv151105487B,2016arXiv160106770B,2016arXiv160202754M}. 

Similar calculations for the formation of dust grains in the ejecta of supernovae type Ia rely on the predictions from \citet{2011ApJ...736...45N} based on theoretical models with values ranging from 3 $\times$ 10$^{-4}$ to 0.2 M$_{\odot}$ depending on the formation of molecules such as CO and SiO. For this study, we rely on the upper ejected mass estimate (0.2 M$_{\odot}$) and assume a SN Ia fraction of $\sim$25$\%$ \citep{2014AJ....148...13R}. Given that the presence of a cool (25-50 K) dust reservoir of mass $\geq$ 0.07 M$_{\odot}$ has not been observed in the ejecta of SNe Ia to date (e.g., \citealt{2012MNRAS.420.3557G}), the true dust yields will likely be lower for supernova type Ia. The SN rates for different supernova types in NGC\,185 and NGC\,205 were already discussed in Section \ref{DustLifetime}. Due to the production of both carbonaceous and silicate grains in supernova remnants, we calculate their dust production within the time interval that is determined by the shortest and longest dust survival time for both grain types. We, furthermore, assume that the contribution from SNe to the dust mass budget in NGC\,147 is negligible due to the absence of any star formation activity during the last $\sim$ 1 Gyr \citep{1997AJ....113.1001H}. 

Another important source of dust production in the interstellar medium of galaxies might be attributed to luminous blue variables (LBVs) with an average dust mass-loss rate of 2.5 $\times$ 10$^{-6}$ M$_{\odot}$ yr$^{-1}$ \citep{2014A&A...569A..80G}. The identification of LBVs is, however, difficult due to the variability of the star's luminosity. With only six confirmed classifications of LBVs in the LMC \citep{1994PASP..106.1025H} and the lack of any LBV identified in any of the three dwarf spheroidals, we do not consider the contribution of dust mass lost by LBVs.

To compute the total dust mass injected in the ISM, we sum the lower and upper mass estimates for the dust mass contributions from AGB stars and supernovae over a period similar to the dust survival time scale (see last part of Table \ref{dustmassloss}). For NGC\,147, the dust mass injected by AGB stars ($\sim$ 189\,M$_{\odot}$) is consistent within the error bars with the upper dust mass limit derived from \textit{Herschel} observations in this paper ($\leq$\,128$^{+124}_{-68}$ M$_{\odot}$). The highest values for the total injected dust mass in NGC\,185 (124 M$_{\odot}$) and NGC\,205 (1026 M$_{\odot}$) are, however, more than one order of magnitude lower compared to the observed dust content in NGC\,185 (5.1 $\times$ 10$^{3}$ M$_{\odot}$) and NGC\,205 (1.1-1.8 $\times$ 10$^{4}$ M$_{\odot}$). The incompatibility between the predicted and observed dust content is thought to be even more extreme considering that mass returned by AGB stars will be distributed across the entire stellar disk while the current dust reservoir appears restricted to the central few 100 pc regions in NGC\,185 and NGC\,205. 

While AGBs are the most important dust resources in our own Galaxy \citep{1998ApJ...501..643D,2008A&A...479..453Z}, AGB stars and SNe seem to be equally important dust producers in the Large Magellanic Clouds \citep{2009MNRAS.396..918M}. Based on our above predictions, we infer that the dust mass produced in supernovae might have a higher contribution to the current dust reservoir in NGC\,185 than the dust mass returned by evolved stars\footnote{Note that the supernova rate in NGC\,185 might be 4 times lower \citep{2012MNRAS.419.3159M}, which would make AGB stars and supernovae equally efficient dust producers.}, while AGB stars and supernova have a similar contribution to the dust mass budget in NGC\,205. The latter inference is, however, strongly dependant on our assumptions of the dust production rate in SNe and AGB stars and the dust survival times of carbonaceous and silicate grains. Especially, the production and destruction of dust by SNe requires more observational studies to confirm the efficiency of dust production in SNe that was predicted from theoretical models.  

\subsection{Implications for dust evolution models}
\label{DustImplications}

The inconsistency between the dust mass predictions and observations suggests that the dust mass production rate has been underestimated and/or the dust survival time is longer compared to the estimated values. The interstellar medium in those dwarf spheroidal galaxies might provide an environment less hostile and retain dust grains for a longer time before being destroyed in grain shattering or sputtering processes. To produce the observed dust content in the dwarf spheroidals NGC\,185 and NGC\,205, we would require a continuous ejection of dusty material at a pace of 1.65 $\times$ 10$^{-6}$ M$_{\odot}$ yr$^{-1}$ and 2.93 $\times$ 10$^{-6}$ M$_{\odot}$ yr$^{-1}$, respectively, for 3 and 4-6 Gyr without any dust destruction\footnote{The dust mass injection rates are determined based on the upper limit of the total dust mass returned by AGB stars and SNe in NGC\,185 (124\,M$_{\odot}$) and NGC\,205 (1026\,M$_{\odot}$) over an average dust survival time for carbon and silicate grains of 75\,Myr and 350\,Myr, respectively.}. While a lifetime of 3\,Gyr for the dust in NGC\,185 could be compatible with the latest predictions for the dust survival time of 2-3 Gyr by \citet{2015ApJ...803....7S}, a dust lifetime of $\sim$ 4-6\,Gyr for NGC\,205 is significantly higher compared to the values predicted by current dust evolution models.

The dust reservoirs in NGC\,185 and NGC\,205 could, alternatively, be of external origin. We, however, argue that the dust reservoir is produced within the galaxy, given that the dusty clouds are located in the very central regions of NGC\,185 and NGC\,205 and spatially correlate with the gaseous material returned by evolved stars. A similar inconsistency between the dust mass returned by stellar sources and the observed dust-to-stellar mass ratio is observed in more massive elliptical galaxies with hardly any ongoing star formation activity \citep{2016MNRAS.456.2221D}. For more massive ellipticals, a scenario where the dusty material has been accreted through galaxy interaction seems more plausible (e.g., \citealt{2012ApJ...748..123S,2013A&A...552A...8D}).

Since AGBs and SNe can not account for the production of the current dust reservoir in NGC\,185 and NGC\,205 (during the estimated dust survival times for these galaxies) and the origin of grain species is considered to be internal, we argue that a significant fraction of the dust mass in dwarf spheroidal galaxies results from grain growth by accretion or coagulation. The importance of grain growth has been invoked for many different types of galaxies (e.g., \citealt{1993A&A...280..617O,2003A&A...398..551S,2009A&A...506..745P,2011MNRAS.417.1510D,2012A&A...548A..61K,2012MNRAS.423...26M,2013EP&S...65..213A,2014MNRAS.444..797M,2014A&A...563A..31R}, and suggest that our current dust evolution models should be updated to include the contribution from material accreting onto pre-existing grains in the densest phases of the ISM.

With the dust mass surface density peaking in regions with the highest gas densities (\citealt{2012MNRAS.423.2359D}, and see Figure \ref{Ima_comb}, right panel, in this paper), our scenario appears at least consistent with efficient metal accretion processes on to the surface of pre-existing grains in the cold neutral medium. Based on semi-analytic chemical evolution models, \citet{2016MNRAS.457.1842S} show that the ISM density is the dominant factor in setting the efficiency of grain growth. In SBS\,0335-052, they attribute 85$\%$ of the current dust mass to grain growth in the dense phase with molecular gas densities of $n_{\text{mol}}$ $\sim$ 1500 cm$^{-3}$, while grain growth can account for only 20$\%$ of the dust mass observed in I\,Zw\,18 with molecular gas densities of only $n_{\text{mol}}$ $\sim$ 100 cm$^{-3}$. Based on the high gas densities derived for NGC\,185 based on PDR models ($n_{\text{H}}$ = 10$^{4}$ cm$^{-3}$, \citealt{DeLooze_paper2}), we believe that grain growth in the densest regions of the ISM provides a plausible explanation for the observed dust content in NGC\,185 (and NGC\,205). 

\subsection{Possible caveats}
\label{PosCaveats}

We acknowledge that the latter predictions for dust mass production are uncertain, and depend on the dust lifetimes and dust injection rates. The dust mass-loss rates for oxygen-rich AGBs stars might alter with metallicity (e.g., \citealt{1998A&A...336..925W,2004MNRAS.355.1348M}), while the dust mass-loss rates for carbon-rich stars seem to depend less on metallicity (e.g., \citealt{2007MNRAS.376..313G,2007MNRAS.382.1889M}). \citet{2015ApJ...800...51B}, however, do not find any dependence of dust mass production rate on metallicity in their sample of nearby dwarf galaxies (including NGC\,147 and NGC\,185). The predictions of mass-loss rates also critically depend on the condensation temperature of dust with a typical value of 1000 K \citep{2006A&A...448..181G,2008ApJ...688L...9G,2009MNRAS.396..918M}. For higher carbon-to-oxygen ratios, we might find higher condensation temperatures for graphite grains, resulting in mass-loss rates higher by a factor of $\sim$ 2.4 \citep{2009MNRAS.396..918M}. 
 
Predictions of the dust mass production rates might, furthermore, be affected by variations in the initial mass function. More specifically, \citet{2016MNRAS.455.4183V} show that the stellar yields are a factor of two higher for a \citet{2001MNRAS.322..231K} and \citet{2003PASP..115..763C} IMF compared to a \citet{1955ApJ...121..161S} IMF (mostly due to differences in the oxygen yields), while the stellar yields for a \citet{1993MNRAS.262..545K} IMF are even lower compared to the latter IMF by a factor of roughly two\footnote{The \citet{1955ApJ...121..161S} IMF has a power law index $\alpha$=2.35 across the entire stellar mass range, while the \citet{2001MNRAS.322..231K} and \citet{2003PASP..115..763C} IMFs assumes broken power laws with index $\alpha$=2.35 above stellar masses of 0.5 M$_{\odot}$ and 1 M$_{\odot}$, respectively. In the low stellar mass regime, the \citet{2001MNRAS.322..231K} IMF assumes a power law index of 0.3 and 1.3 for the stellar mass ranges $<$0.08\,M$_{\odot}$ and 0.08\,M$_{\odot}$$<$M$<$0.5\,M$_{\odot}$, respectively. The \citet{2003PASP..115..763C} IMF for individual stars closely follows the \citet{2001MNRAS.322..231K} at these low stellar masses, and is described by an analytic function dependent on stellar mass. The \citet{1993MNRAS.262..545K} IMF has a steeper power law ($\alpha$=2.7) for stellar masses M$>$1M$_{\odot}$, while $\alpha$=2.2 and 0.70$<$$\alpha$$<$1.85 in the mass ranges 0.5$<$M$<$1M$_{\odot}$ and 0.08$<$M$<$0.5M$_{\odot}$.}. Variations in the assumed upper mass limit (100\,M$_{\odot}$ instead of 40\,M$_{\odot}$) can, furthermore, result in higher stellar mass yields by factors of two and four for a \citet{1955ApJ...121..161S} and \citet{2003PASP..115..763C} IMF, respectively.

\citet{2013ApJ...771...29G} showed that the low mass IMF slopes in metal-poor dwarf galaxies become shallower in their studied stellar mass range of 0.52-0.77 M$_{\odot}$, resulting in a more bottom-light initial mass function and, thus, higher stellar yields. For NGC\,185, we estimate an IMF slope of $\alpha$ = 2.0 in the stellar mass range 0.52-0.77 M$_{\odot}$ (compared to $\alpha$ = 2.35 for a \citealt{1955ApJ...121..161S}, \citealt{2001MNRAS.322..231K} and \citealt{2003ApJ...598.1076K} IMF) based on the observed velocity dispersion ($\sim$ 24 km s$^{-1}$, \citealt{2010ApJ...711..361G}) and metallicity ([Fe/H] $\sim$ -1.0, \citealt{2015ApJ...811..114G}). Similarly, the velocity dispersion ($\sigma$ $\sim$ 20-30 km s$^{-1}$, \citealt{2006MNRAS.369.1321D}) and metallicity ([Fe/H] $\sim$ -0.9, \citealt{2005MNRAS.356..979M}) in NGC\,205 would imply an IMF slope close to $\alpha$ = 2.0 following the scaling relations for the IMF slope reported by \citet{2013ApJ...771...29G}. 

More studies on the IMF shape in dwarf spheroidal galaxies over larger stellar mass ranges are required to make accurate predictions about the effect of the IMF slope on stellar yields in these galaxies. For the dust mass injection rates adopted from \citet{2012ApJ...753...71R}, no particular assumptions about the IMF were applied since the mass loss rates are derived from a direct fit of a set of radiative transfer models to the optical to mid-infrared photometry of $\sim$30,000 AGB and RSG stars in the LMC. Other than the shape of the IMF, the small physical size of the central star-forming region (a few 100\,pc) is likely not fully sampling the stellar mass function which might affect the estimation of the star formation rate and/or supernova rate.

By scaling the dust mass production rates derived for the LMC by \citet{2012ApJ...753...71R} to the observed population of AGB stars in the dSphs from the Local Group, we assume that the stellar mass distribution of evolved stars in those dwarfs is similar to the LMC evolved stellar population. \citet{2005AJ....129.2217B}, however, showed that the RGB and faint-AGB is skewed to redder colours for NGC\,205 (while this is not the case for NGC\,185 or NGC\,147), which might indicate a narrower range in stellar abundances or age for the AGB stars in NGC\,205. The near-infrared observations used by \citet{2005AJ....130.2087D} might, furthermore, miss detecting some of the fainter AGB stars, which would imply that we, currently, underestimate the stellar yields in these Local Group dSphs. It is, however, unlikely that a change in the slope of the IMF, or an incomplete AGB catalogue could make up for the order of magnitude difference between the dust mass produced by evolved stars and the observed dust content.

\begin{table}
\caption{Overview of the dust mass loss rates for evolved stars obtained by scaling the dust mass injection rates for C-rich AGB, O-rich ABG and RGB populations in the LMC to the observed AGB population in NGC\,147, NGC\,185 and NGC\,205 (top part). The estimated dust mass lost by AGB stars and different types of supernova remnants for the three dSphs of Andromeda during typical dust survival times are presented in the second part of the Table.}
\label{dustmassloss}
\centering
\begin{tabular}{lccc}
\hline 
Population & \multicolumn{3}{|c}{Dust mass loss rate $\dot{M_{\text{d}}}$ [10$^{-8}$ M$_{\odot}$ yr$^{-1}$]} \\
\hline
 & NGC\,147 & NGC\,185 & NGC\,205 \\
\hline
C-rich stars & 12.2 & 19.3 & 72.5 \\
O-rich stars & 4.9 & 7.8 & 29.2 \\
RSGs & 1.8 & 2.8 & 10.6 \\
\hline
Total AGB population & 18.9 & 29.9 & 112.3 \\
\hline \hline  
Population & \multicolumn{3}{|c}{Dust mass loss [M$_{\odot}$]} \\
\hline
 & NGC\,147 & NGC\,185 & NGC\,205 \\
 \hline
C-rich AGB stars & 122 & 4-10 & 65-207 \\
 O-rich AGB stars + RSGs & 67 & 4-11 & 71-228 \\
 SNe II+Ib & / & 11-55 & 49-316 \\
 SNe Ia & / & 10-48 & 43-275 \\
\hline 
Total & 189 & 29-124 & 228-1026 \\ 
\hline 
\end{tabular}
\end{table}

\section{Conclusions}
\label{Conclusions.sec}
In this paper, we analyse the dust reservoir in the dwarf spheroidal galaxies NGC\,147 and NGC\,185 based on \textit{Herschel} PACS and SPIRE observations. In a companion paper \citep{DeLooze_paper2}, we revise the gas content of the dSphs in the Local Group and discuss their gas content, ISM properties and gas-to-dust ratio in view of galaxy evolution scenarios for the dwarf spheroidal galaxy population.
\begin{itemize}
\item Similar to the non-detections of any gas in the ISM of NGC\,147, no dust emission is detected from NGC\,147 in the SPIRE maps, which puts a strong upper limit on its dust content ($M_{\text{d}}$ $\leq$ 128$^{+124}_{-68}$ M$_{\odot}$).
The SPIRE dust maps of NGC\,185 show an extended cold dust component in the East of the galaxy, co-spatial with the H{\sc{i}} distribution, that was not yet detected in far-infrared wavebands. From a full dust model SED fitting method, we derive an average scaling factor of the radiation field $G_{0}$ $\sim$ 1.5-2.1 and a total dust mass $M_{\text{d}}$ = 5.1 $\times$ 10$^{3}$ M$_{\odot}$, which is factors of 2-3 higher compared to the dust reservoirs inferred from \textit{ISO} and \textit{Spitzer} observations. This confirms that observational constraints at wavelengths longward of 160-200\,$\mu$m are a prerequisite to derive reliable dust masses. 

\item The best fitting dust SED model for NGC\,185 that is optimised to reproduce the \textit{Herschel} observations, overestimates the observed flux at millimetre wavelengths reported in the literature, which can, most likely, be attributed to the small beam throw used for the IRAM\,1.2\,mm observations. We, furthermore, conclude that the observed IRAC\,8\,$\mu$m emission is consistent with a dust model with Galactic PAH abundance, and that NGC\,185 does not show any evidence for excess emission in the submillimeter wavelength domain, despite its low metal fraction (0.36 Z$_{\odot}$).

\item We compare the dust mass content observed in all three dSph companions of the Andromeda galaxy to predictions of the mass returned by AGB stars and supernova remnants. Making reasonable assumptions about the dust mass-loss rates from asymptotic giant branch stars and supernova of different types, we compute the expected dust mass returned to the ISM over a time period comparable to the dust survival time in NGC\,147 (189\,M$_{\odot}$), NGC\,185 (29-124\,M$_{\odot}$), and NGC\,205 (228-1026\,M$_{\odot}$). The dust mass injected by evolved stars is consistent with the upper dust mass limit ($<$ 128$^{+124}_{-68}$ M$_{\odot}$) in NGC\,147. For the other two dSph companions, the latter predictions are more than one order of magnitude lower than the observed dust content in NGC\,185 ($M_{\text{d}}$ = 5.1 $\times$ 10$^{3}$ M$_{\odot}$) and NGC\,205 ($M_{\text{d}}$ = 1.1-1.8 $\times$ 10$^{4}$ M$_{\odot}$). We, therefore, argue that other dust resources (e.g., grain growth through accretion of elements from the gas phase) have an important contribution to the interstellar dust production. Longer dust survival times ($\sim$ 3-6 Gyr) could also aid to bridge the gap between model predictions and observations, but are unlikely to account for the current interstellar dust content.

\item In the future, we need high-spatial resolution near- and mid-infrared observations to constrain the dominant dust production sources in a larger sample of galaxies to assess the importance of dust destruction and grain growth in the ISM. In a first step, the infrared observations for 50 local dwarf galaxies from DUSTiNGS (DUST in Nearby Galaxies with Spitzer, \citealt{2015ApJS..216...10B}) will provide additional constraints on the dust mass production rate for evolved stars in NGC\,147 and NGC\,185, and other nearby dwarf galaxies. The advent of the James Webb Space telescope will, in the future, open up the pathway for spatially resolved stellar population studies in a large sample of galaxies. 
\end{itemize}

\section*{Acknowledgments}
The authors would like to thank Anthony Jones and Marla Geha for interesting discussions that have helped to improve this paper.
IDL gratefully acknowledge the support of the Science and Technology Facilities Council (STFC) and the Flemish Fund for Scientific Research (FWO-Vlaanderen).
PACS has been developed by a consortium of institutes
led by MPE (Germany) and including UVIE
(Austria); KU Leuven, CSL, IMEC (Belgium);
CEA, LAM (France); MPIA (Germany); INAFIFSI/
OAA/OAP/OAT, LENS, SISSA (Italy);
IAC (Spain). This development has been supported
by the funding agencies BMVIT (Austria),
ESA-PRODEX (Belgium), CEA/CNES (France),
DLR (Germany), ASI/INAF (Italy), and CICYT/
MCYT (Spain). SPIRE has been developed
by a consortium of institutes led by Cardiff
University (UK) and including Univ. Lethbridge
(Canada); NAOC (China); CEA, LAM
(France); IFSI, Univ. Padua (Italy); IAC (Spain);
Stockholm Observatory (Sweden); Imperial College
London, RAL, UCL-MSSL, UKATC, Univ.
Sussex (UK); and Caltech, JPL, NHSC, Univ.
Colorado (USA). This development has been
supported by national funding agencies: CSA
(Canada); NAOC (China); CEA, CNES, CNRS
(France); ASI (Italy); MCINN (Spain); SNSB
(Sweden); STFC and UKSA (UK); and NASA
(USA). 
This research has made use of the NASA/IPAC Extragalactic Database (NED) which is operated by the Jet Propulsion Laboratory, California Institute of Technology, under contract with the National Aeronautics and Space Administration.


\label{lastpage}

\end{document}